\documentclass[trackchanges]{aastex701}

\usepackage[noend]{algpseudocode}
\usepackage{newfloat}
\usepackage{caption}

\DeclareFloatingEnvironment[
    fileext=loa,
    listname=List of Algorithms,
    name=Algorithm
]{algorithm}

\usepackage{multirow}

\newcommand{\kms}{km~s$^{-1}$}
\newcommand{\kmsspace}{km~s$^{-1}$ }

\newcommand{\vsw}{$V_\mathrm{sw}$ }
\newcommand{\rs}{$r_\odot$}
\newcommand{\rsspace}{$r_\odot$ }
\newcommand{\donkilon}{$\mathrm{lon}_{\mathrm{ DONKI}} $}
\newcommand{\donkilat}{$\mathrm{lat}_{\mathrm{ DONKI}} $}

\received{Feb 6}
\revised{Mar 20}
\accepted{Apr 9}
\submitjournal{ApJ}

\begin{document}

\title{Towards a Fully Automated Pipeline for Short-Term Forecasting of In Situ Coronal Mass Ejection Magnetic Field Structure}

% Towards a Fully Automated End-to-End Pipeline for Short-Term CME Magnetic Field Forecasting
% ARCANE - Combining Automatic Flux Rope Fitting and Real-Time ICME Detection} %for Operational Short-Term Forecasting}

%%%%%%%%%%%%%%%%%%%%%%%%%%%%%%%%%%%%%%%%%%%%%%%
%  AUTHORS
%%%%%%%%%%%%%%%%%%%%%%%%%%%%%%%%%%%%%%%%%%%%%%%

\author[0000-0002-2559-2669]{Hannah T. Rüdisser}
\affiliation{Austrian Space Weather Office, GeoSphere Austria, Graz, Austria}
\affiliation{Institute of Physics, University of Graz, Graz, Austria}
\email{hannah@ruedisser.at}

\correspondingauthor{Hannah T. Rüdisser}
\email{hannah@ruedisser.at}

\author[0000-0001-9992-8471]{Emma E. Davies}
\affiliation{Austrian Space Weather Office, GeoSphere Austria, Graz, Austria}
\email{emma.davies@geosphere.at}

\author[0000-0003-1516-5441]{Ute V. Amerstorfer}
\affiliation{Austrian Space Weather Office, GeoSphere Austria, Graz, Austria}
\email{ute.amerstorfer@geosphere.at}

\author[0000-0001-6868-4152]{Christian Möstl}
\affiliation{Austrian Space Weather Office, GeoSphere Austria, Graz, Austria}
\email{christian.moestl@geosphere.at}

\author[0009-0004-8761-3789]{Eva Weiler}
\affiliation{Austrian Space Weather Office, GeoSphere Austria, Graz, Austria}
\affiliation{Institute of Physics, University of Graz, Graz, Austria}
\email{eva.weiler@geosphere.at}

\author[0000-0002-6273-4320]{Andreas J. Weiss}
\affiliation{Goddard Planetary Heliophysics Institute, University of Maryland, Baltimore County, Baltimore, MD 21250, USA}
\affiliation{Heliophysics Science Division, NASA Goddard Space Flight Center, Greenbelt, MD, USA}
\email{ajefweiss@gmail.com}

\author[0000-0001-5387-9512]{Justin Le Louëdec}
\affiliation{Austrian Space Weather Office, GeoSphere Austria, Graz, Austria}
\email{justin.lelouedec@geosphere.at}

\author[0000-0002-6362-5054]{Martin A. Reiss}
\affiliation{Community Coordinated Modeling Center, NASA Goddard Space Flight Center, 8800 Greenbelt Rd., Greenbelt, MD 20771, USA}
\affiliation{Universities Space Research Association, Washington, DC, USA}
\email{martin.a.reiss@outlook.com}

\author[0000-0003-4955-9924]{Gautier Nguyen}
\affiliation{DPHY, ONERA, Université de Toulouse, F-31000, Toulouse, France}
\email{gautier.nguyenc@onera.fr}

%%%%%%%%%%%%%%%%%%%%%%%%%%%%%%%%%%%%%%%%%%%%%%%
%  ABSTRACT
%%%%%%%%%%%%%%%%%%%%%%%%%%%%%%%%%%%%%%%%%%%%%%%

\begin{abstract}

We present the automated pipeline NEXUS (Near-real-time Event detection and eXtrapolation using a Unified Space weather pipeline) for operational short-term forecasting of coronal mass ejection (CME) magnetic field structure at L1, coupling arrival time prediction, in situ detection, and iterative flux rope reconstruction, following near-real-time remote-sensing CME identification. Triggered by entries in the CCMC DONKI database, NEXUS first applies the drag-based ELEvo model to determine whether an Earth impact is expected and estimates arrival time. This defines a temporal window for detecting the magnetic obstacle (MO) in real-time L1 in situ solar wind data, using the deep learning model ARCANE. Upon MO onset, iterative reconstructions with the semi-empirical flux rope model 3DCORE are performed, using a Monte Carlo fitting scheme, producing continuously updated forecasts of the remaining magnetic field profile.

We evaluate NEXUS using 3870 archived DONKI entries and NOAA real-time in situ data from 2013-2025, assessing forecast performance at different stages of MO observation. For 61 events with an associated counterpart in the ICMECAT catalog, forecasts based on initial MO data already achieve performance comparable to full-event reconstructions. Typical errors are $\sim5$~hours in timing of magnetic field extrema and $\sim10$~nT in field strength metrics, with limited systematic improvement as more of the event is observed. Results show substantial event-to-event variability and systematic underestimation of extrema, indicating deviations from ideal flux rope assumptions. These findings demonstrate the potential of fully autonomous real-time forecasting while highlighting limitations imposed by event complexity and model constraints.

\end{abstract}

%%%%%%%%%%%%%%%%%%%%%%%%%%%%%%%%%%%%%%%%%%%%%%%
%  KEYWORDS
%%%%%%%%%%%%%%%%%%%%%%%%%%%%%%%%%%%%%%%%%%%%%%%



%%%%%%%%%%%%%%%%%%%%%%%%%%%%%%%%%%%%%%%%%%%%%%%
%  INTRODUCTION
%%%%%%%%%%%%%%%%%%%%%%%%%%%%%%%%%%%%%%%%%%%%%%%

\section{Introduction} \label{sec:intro}

Coronal mass ejections (CMEs) are the primary drivers of significant disturbances in the Earth's magnetosphere, posing a substantial risk to critical infrastructure, including satellite operations, communication networks, GNSS-based navigation and terrestrial power grids \citep[e.g.][]{eastwood_2017_economicimpactspace, kilpua_2019_forecastingstructureorientation, vourlidas_2019_predictinggeoeffectiveproperties, owens_2020_valuecmearrival, hapgood_2021_developmentspaceweather}. Mitigating these risks requires timely and reliable forecasts that provide actionable information.

Despite decades of research and a growing fleet of solar and heliospheric space missions, the forecasting of CME arrival and geomagnetic effect remains challenging. Uncertainties arise from limited observational constraints on geometry and internal structure as well as complex propagation through background solar wind. Additionally, there is the need for computationally efficient methods that can operate in real-time \citep[e.g.][]{temmer_2023_cmepropagationheliosphere}. An inaccurate forecast can either lead to unnecessary and costly mitigation efforts or, conversely, to a failure to protect vulnerable systems \citep[e.g.][]{eastwood_2017_economicimpactspace, oughton_2017_quantifyingdailyeconomic}.

Our ability to predict CME arrival and geomagnetic effect is intrinsically linked to our understanding of CME structure and evolution. Our understanding of CME structure is built upon two primary categories of observations: remote-sensing and in situ measurements. Remote-sensing observations, such as coronagraph images, established the classic ``three-part" structure, consisting of a bright leading edge, dark cavity, and bright core; the magnetic structure, however, is not directly observed. In situ observations, in contrast, provide direct magnetic field and plasma measurements, but typically only at a single point in space, probing the CME directly as it sweeps over the spacecraft. In interplanetary space, CMEs (often referred to as ICMEs) commonly exhibit a shock, a turbulent sheath, and a magnetic obstacle (MO). A subset of MOs are classified as magnetic clouds, characterized by enhanced magnetic field strength, smooth field rotation, low proton temperature, and low plasma beta \citep[e.g.][]{bothmer_1998_structureoriginmagnetic,cane_2003_interplanetarycoronalmass,zurbuchen_2006_insitusolarwind}. These signatures are often interpreted as evidence for a flux rope topology, frequently modeled using idealized solutions such as the Lundquist or Gold-Hoyle configurations \citep{lundquist_1950_magnetohydrostaticfields,gold_1960_originsolarflares}. This coherent structure is also what can sustain extended intervals of southward $B_Z$ and thereby drive intense geomagnetic activity, making CME geomagnetic effect forecasting inseparable from forecasting the CME's internal magnetic structure \citep[e.g.][]{vourlidas_2019_predictinggeoeffectiveproperties,temmer_2023_cmepropagationheliosphere}.

For many years, the standard model of a CME in interplanetary space was that of a highly twisted circular cross-section flux rope and became the foundation for many analytical and numerical models \citep[e.g.][]{lepping_1990_magneticfieldstructure,hidalgo_2012_globalmagnetictopology,isavnin_2016_friednovelthreedimensional,mostl_2018_forwardmodelingcoronal,rouillard_2020_modelingearlyevolution,nieves-chinchilla_2020_analysisinternalstructure}. However, decades of observations show that this interpretation is not universally adequate and forced a revision of this picture. \cite{gosling_1990_coronalmassejections} established that only around 30\% of ICMEs at 1~au exhibit the classic, large-scale coherent field rotations characteristic of magnetic flux ropes, with this fraction further decreasing during solar maximum \citep{cane_2003_interplanetarycoronalmass}. In situ multi-spacecraft studies further demonstrate that apparent coherence in single-spacecraft time series can break down over larger spacecraft separations due to deformation, interaction, erosion, and other propagation effects \citep[e.g.][]{scolini_2022_causesconsequencesmagnetic, lugaz_2018_spatialcoherencemagnetic,salman_2020_radialevolutioncoronal, davies_2022_multispacecraftobservationsevolutiona}. Consequently, flux rope identification and parameter inference depend strongly on model assumptions and on how closely the observed event resembles the idealized topology \citep{riley_2004_fittingfluxropes,lynch_2022_utilityfluxrope}.

Recent work by \cite{al-haddad_2025_magneticfieldstructure} emphasizes a significantly more complex and less idealized view of CME structure. The key features of this updated CME representation are a kinked central axis, which is often distorted, a non-uniform cross-section that evolves as it interacts with the ambient solar wind, mixed magnetic topology, containing a mixture of open and closed magnetic field lines throughout their structure, and variable twist, where the degree of magnetic twist within the structure is assumed to be relatively small. These complexities are expected outcomes of propagation through structured solar wind, interaction with high-speed streams and corotating interaction regions \citep[e.g.][]{webb_2012_coronalmassejections, palmerio_2022_cmeevolutionstructured}, rotation and deflection \citep[e.g.][]{wang_2004_deflectioncoronalmass, kay_2015_globaltrendscme}, magnetic erosion \citep[e.g.][]{dasso_2007_progressivetransformationflux, ruffenach_2015_statisticalstudymagnetic}, and CME-CME interactions \citep[e.g.][]{lugaz_2017_interactionsuccessivecoronal, scolini_2020_cmecmeinteractionssources} that can produce complex ejecta in which individual flux rope boundaries are ambiguous \citep[e.g.][]{zurbuchen_2006_insitusolarwind,kilpua_2017_coronalmassejections,good_2018_correlationicmemagnetica,lugaz_2018_spatialcoherencemagnetic, kay_2026_collectioncollationcomparison}. In addition, purely kinematic effects such as ``pancaking" can distort cross-sections from circular to flattened or elongated shapes \citep[e.g.][]{savani_2011_evolutioncoronalmass, davies_2021_situmultispacecraftremote}. While more flexible models are being developed to represent such distortions more realistically \citep[e.g.][]{weiss_2024_distortedmagneticflux}, operational forecasting remains constrained by sparse sampling: the majority of CMEs are still effectively measured at only a single in situ point \citep{lugaz_2024_neednearearthmultispacecraft}, forcing simplified assumptions in both research and operations.

Forecasting CME arrival times has long reached a plateaued state. Independent of the used model (drag-based, MHD or machine learning), forecasts typically do not reach mean absolute errors in arrival time below 13 hours, as shown with model comparisons on the CCMC Scoreboard \citep{riley_2018_forecastingarrivaltime} and more recent benchmarks \citep[e.g.][]{kay_2024_updatingmeasurescme}. Continued improvements in remote-sensing constraints have not yet produced systematic gains in performance \citep[e.g.][]{vourlidas_2019_predictinggeoeffectiveproperties, owens_2020_valuecmearrival}. Current studies and efforts now increasingly target the forecasting of the magnetic structure, recognizing it as the main bottleneck for accurate space weather prediction \citep[e.g.][]{kubicka_2016_predictiongeomagneticstorm, laker_2024_usingsolarorbiter, lugaz_2024_neednearearthmultispacecraft, lugaz_2025_needsubl1space, davies_2025_realtimepredictiongeomagneticb}.

Two principal pathways have emerged for predicting $B_Z$. First, Sun-to-Earth modeling, which attempts to infer the CMEs magnetic configuration from remote-sensing observations and propagate it through the heliosphere using physics-based or semi-empirical models. Recent examples include coupled modeling frameworks such as OSPREI \citep{kay_2022_ospreicoupledapproach}, which combines coronal deflection/rotation, heliospheric propagation and deformation, and synthetic in situ profile generation to predict arrival time, impact geometry, and internal magnetic properties. Such approaches are physically comprehensive but often limited by uncertain boundary constraints and computational cost. Second, upstream monitoring leverages earlier in situ measurements to predict the magnetic field profile later at L1 \citep[e.g.][]{kubicka_2016_predictiongeomagneticstorm, laker_2024_usingsolarorbiter, moreno_2025_xcmesitufluxrope, mao_2025_inferringmagneticfield, weiler_2025_firstobservationsgeomagnetic, davies_2025_realtimepredictiongeomagneticb}, providing short term warning that can be particularly valuable operationally \citep[e.g.][]{palmerio_2025_monitoringsolarwind}. This approach exploits the inherent coherence of flux ropes, making it possible to extrapolate the magnetic field evolution from early observations. At present, however, its applicability is constrained by the scarcity of operational upstream monitors and the limited frequency of favorable spacecraft configurations \citep{lugaz_2024_neednearearthmultispacecraft, palmerio_2025_monitoringsolarwind}.

A complementary third pathway, and the focus of this study is short-term in situ forecasting. Once the ICME magnetic obstacle begins to be observed at a monitoring point, the already observed fraction of the structure is used to infer and predict the remaining magnetic profile and its geoeffectiveness  \citep[e.g.][]{ chen_1997_predictingsolarwind, telloni_2019_detectioncoronalmass, reiss_2021_machinelearningpredicting, pal_2024_automaticdetectionlargescale}. However, a major obstacle to operational deployment is that many established analysis tools remain case-specific and require a human-in-the-loop. Traditional flux rope fitting typically requires manual selection of intervals, initial parameter choices and assumptions about handedness or geometry \citep[e.g.][]{riley_2004_fittingfluxropes, lynch_2022_utilityfluxrope}. In contrast, real-time forecasting demands autonomous pipelines that can process continuous data streams, make decisions without human intervention and update predictions as new measurements become available.

In this work, we present the first fully automated pipeline that connects real-time remote-sensing and in situ data streams to produce short-term CME magnetic field forecasts. The NEXUS (Near-real-time Event detection and eXtrapolation using a Unified Space weather pipeline) pipeline connects three complementary models that address different stages of CME forecasting: arrival time prediction, in situ event detection, and magnetic field structure reconstruction. Specifically, the system integrates:

\begin{enumerate}
    \item CME arrival time prediction using ELEvo \citep{mostl_2015_strongcoronalchannelling}, a drag-based propagation model;
    \item Automatic MO detection via ARCANE \citep{rudisser_2026_arcaneearlydetectioninterplanetary}, a deep-learning-based model for in situ data;
    \item Iterative flux rope reconstruction with the semi-empirical flux rope model 3DCORE \citep{mostl_2018_forwardmodelingcoronal, weiss_2021_analysiscoronalmass, weiss_2021_multipointanalysiscoronal, rudisser_2024_understandingeffectsspacecraft}, producing continuously updated forecasts of the remaining magnetic profile as the CME sweeps over the observer.
\end{enumerate}

This architecture couples the strengths of three different forecasting models. It uses remote-sensing informed propagation to constrain timing, in situ detection to identify the MO, and a physically interpretable flux rope model to forecast CME magnetic properties on short lead times.

The NEXUS pipeline enables continuous, real-time evaluation of CME propagation and internal structure and provides the first statistical test of the 3DCORE short-term forecast concept. We quantify how forecast quality evolves as progressively larger fractions of the MO are observed and assess performance in predicting operationally relevant quantities, such as the minimum $B_Z$, and their timing. By applying this approach to the entire available dataset, spanning the years 2013-2025, we analyze a total of 3870 DONKI-listed entries. Of these, 406 are predicted by ELEvo to arrive at Earth. Using ARCANE, we detect 102 of these in situ at L1, from which 84 yield successful 3DCORE reconstructions. Together, these constitute the most comprehensive set of 3DCORE flux rope reconstructions to date.

In Section~\ref{sec:data} we describe the real-time data sources used in this study. In Section~\ref{sec:method} we introduce the three different models and describe their integration into the fully automated NEXUS pipeline in Section~\ref{sec:pipeline}. We run this pipeline using a hindcasting approach for the years 2013 to 2025, using archived (near-)real-time data products, while ensuring that only data that would have been available at the corresponding time in a real-time operational setting is used. The results are presented in Section~\ref{sec:results}. The implications of these findings for space weather forecasting are discussed in Section~\ref{sec:discussion}, and the main conclusions and possible future directions are summarized in Section~\ref{sec:summary}.

%%%%%%%%%%%%%%%%%%%%%%%%%%%%%%%%%%%%%%%%%%%%%%%
%  DATA
%%%%%%%%%%%%%%%%%%%%%%%%%%%%%%%%%%%%%%%%%%%%%%%

\section{Data}\label{sec:data}

Our analysis relies on three primary data sources: L1 in situ data, an ICME catalog and remote-sensing data used to initialize models. While individual modules within NEXUS require only a subset of these inputs, testing the complete pipeline necessitates the availability of all three data sources. As a result, the full pipeline can only be evaluated during time periods for which all datasets are simultaneously available.

As described in detail in \cite{rudisser_2026_arcaneearlydetectioninterplanetary}, we utilize the NOAA Real-Time Solar Wind (RTSW) dataset \citep{zwickl_1998_noaarealtimesolarwind}, provided by the Space Weather Prediction Center (NOAA SWPC) and available at \url{https://www.spaceweather.gov/products/real-time-solar-wind}. This dataset spans from 1998 to 2025. Following \cite{rudisser_2026_arcaneearlydetectioninterplanetary} and \cite{nguyen_2025_simultaneousmulticlassdetection}, we select the following parameters: the three Geocentric Solar Magnetic (GSM) components of the interplanetary magnetic field (IMF) ($B_X$, $B_Y$, $B_Z$), their total magnitude ($|B|$), the proton density ($N_p$), the proton temperature ($T_p$), the bulk solar wind speed ($V$), and the plasma beta ($\beta$). To account for short-term data gaps, we use a temporal resolution of 10 minutes, and apply linear interpolation to fill gaps shorter than 6 hours. For further details, the reader is referred to \cite{rudisser_2026_arcaneearlydetectioninterplanetary}.

Following \cite{rudisser_2026_arcaneearlydetectioninterplanetary}, we employ the HELIO4CAST ICME catalog \citep{mostl_2017_modelingobservationssolar, mostl_2020_predictionsitucoronal,moestl_2026_powerlaw_icmecat}, which is continuously updated. In this study, we use version 2.3 \citep{moestl_2025_catalog}, covering the period from February 1995 to July 2025, after restricting the dataset to events observed at L1 (Wind). The catalog contains 551 ICMEs in total, of which 420 include a sheath region. 

The Space Weather Database of Notifications, Knowledge, Information (DONKI), available at \url{https://kauai.ccmc.gsfc.nasa.gov/DONKI/}, is developed at the Community Coordinated Modeling Center (CCMC). Data is provided by the Moon to Mars (M2M) Space Weather Analysis Office and other entities. DONKI offers users near real-time analyses of space weather observations, serving as an operational resource for both research and forecasting applications. Among other parameters, the M2M catalog in DONKI includes estimates of the launch time at 21.5~\rs, longitude ($lon$), latitude ($lat$), half-width ($\lambda$) and initial velocity of events ($V_0$) based on remote sensing data. 

To ensure that the catalog reflects information that would have been available in a near-real-time operational setting, we include only entries for which data was available within 8 hours of the CME reaching 21.5~\rs. Applying this criterion yields 3870 CME IDs between the first record on August 6, 2013 and July 27, 2025. In some cases, multiple reports are associated with the same CME ID. When this occurs, we retain a single entry by prioritizing measurements of the shock front over those of the leading edge and earlier submission times over later ones. While it remains possible that the same physical CME may occasionally receive multiple CME IDs in DONKI, such cases are expected to be rare and difficult to resolve automatically. Of the resulting 3870 entries, 196 are associated with an interplanetary shock (IPS) arrival observed at L1, as reported in DONKI.

%%%%%%%%%%%%%%%%%%%%%%%%%%%%%%%%%%%%%%%%%%%%%%%
%  METHOD
%%%%%%%%%%%%%%%%%%%%%%%%%%%%%%%%%%%%%%%%%%%%%%%

\section{Method}\label{sec:method}

Within our NEXUS real-time pipeline, we employ three complementary models that address the different stages of our forecasting procedure: the ELEvo model for heliospheric propagation and arrival time prediction (Section~\ref{sec:elevo}), the deep-learning-based ARCANE model for early in situ detection of CME substructures (Section~\ref{sec:arcane}), and the semi-empirical flux rope model 3DCORE for iterative reconstruction and short-term forecasting of the CME in situ magnetic field structure. In the following, we describe the setup of each model and assess its performance.

\subsection{ELEvo}\label{sec:elevo}

The Elliptical Evolution (ELEvo) model, first introduced by \cite{mostl_2015_strongcoronalchannelling}, has since been successfully applied in case studies \citep[e.g.][]{weiler_2025_firstobservationsgeomagnetic, davies_2025_realtimepredictiongeomagneticb}. In addition, ELEvo has been employed extensively in its ELEvoHI variant, which extends the basic ELEvo model by further constraining CME kinematics through heliospheric imager observations \citep[e.g.][]{amerstorfer_2021_evaluationcmearrival, amerstorfer_2025_predictingcmearrivals, hinterreiter_2021_whyareelevohia}. It is routinely used to visualize the propagation of CMEs and to predict their arrival times at different locations in the heliosphere. The model assumes an elliptical CME front, which is propagated outward using a simple drag-based model \citep{vrsnak_2013_propagationinterplanetarycoronal}.

ELEvo requires several input parameters: the time the CME is observed at 21.5~\rs, its initial speed ($V_0$) and propagation direction ($lon$, $lat$), its half width ($\lambda$) as well as the ambient solar wind speed ($V_{sw}$) and a drag parameter ($\gamma$) for the drag-based propagation. By varying a subset of these parameters, an ensemble of simulations can be generated to quantify the uncertainty in the predicted arrival times. This approach was extensively discussed in \cite{calogovic_2021_probabilisticdragbasedensemble}.

\begin{figure}
    \begin{center}
    \includegraphics[width=\textwidth]{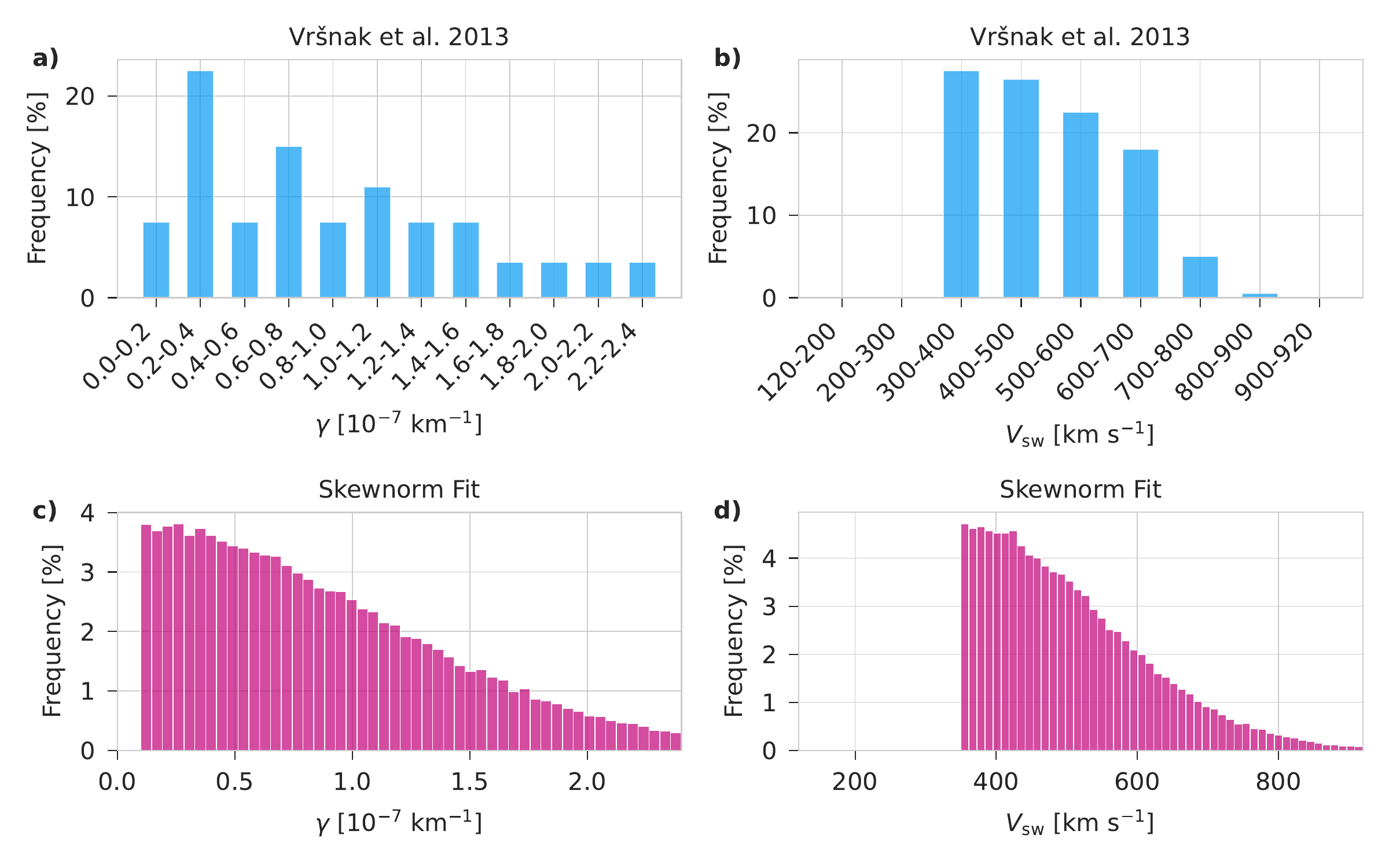}
    \caption{Parameter distributions used in the ELEvo simulations. (a) Histogram of $\gamma$ values based on \cite{vrsnak_2013_propagationinterplanetarycoronal}, where each bar corresponds to the indicated $\gamma$ interval (e.g. 0.0-0.2, 0.2-0.4, etc.) and the bar height denotes the percentage of events in that interval. (b) Histogram of background solar wind speed (\vsw) values from the same source, with the bars representing the labeled speed intervals. (c) Skew-normal fit to the $\gamma$ distribution shown in (a), from which simulation values were sampled. (d) Skew-normal fit to the \vsw distribution in b), used for sampling \vsw values in the simulations.}
    \label{fig:distributions}
    \end{center}
\end{figure}

In this study, we construct such an ensemble by varying the drag parameter $\gamma$, the ambient solar wind speed \vsw and the measured initial CME speed. For $\gamma$, we fit a skewed normal distribution to the values reported in \cite{vrsnak_2013_propagationinterplanetarycoronal} and draw samples from this distribution. A skewed normal distribution is a generalization of the normal distribution that allows for asymmetric (skewed) probability density. In the underlying study, we chose an ensemble size of $10^5$. The histograms extracted from \cite{vrsnak_2013_propagationinterplanetarycoronal}, together with the fitted distributions from which values are sampled, are presented in Figure~\ref{fig:distributions}. To account for uncertainty in the initial CME speed, we follow \cite{kay_2024_collectioncollationcomparison}, who report a typical discrepancy of approximately $115$ \kmsspace between independent CME reconstructions. Accordingly, we sample the initial velocity from a normal distribution centered on the DONKI-reported value, with a standard deviation of $115/2$ \kms.

By analyzing the CME propagation direction and half width, we can determine whether the CME is expected to impact Earth. If an Earth impact is predicted, the ensemble propagation provides a distribution of arrival times from which we compute the mean expected arrival time $t_e$ and its standard deviation $\sigma_{t}$. To define an expected arrival window, we determine  $t_e \pm 2\sigma_{te}$, which corresponds to approximately a 95\% confidence interval under the assumption of a normal distribution.

\begin{figure}
    \begin{center}
    \includegraphics[width=\textwidth]{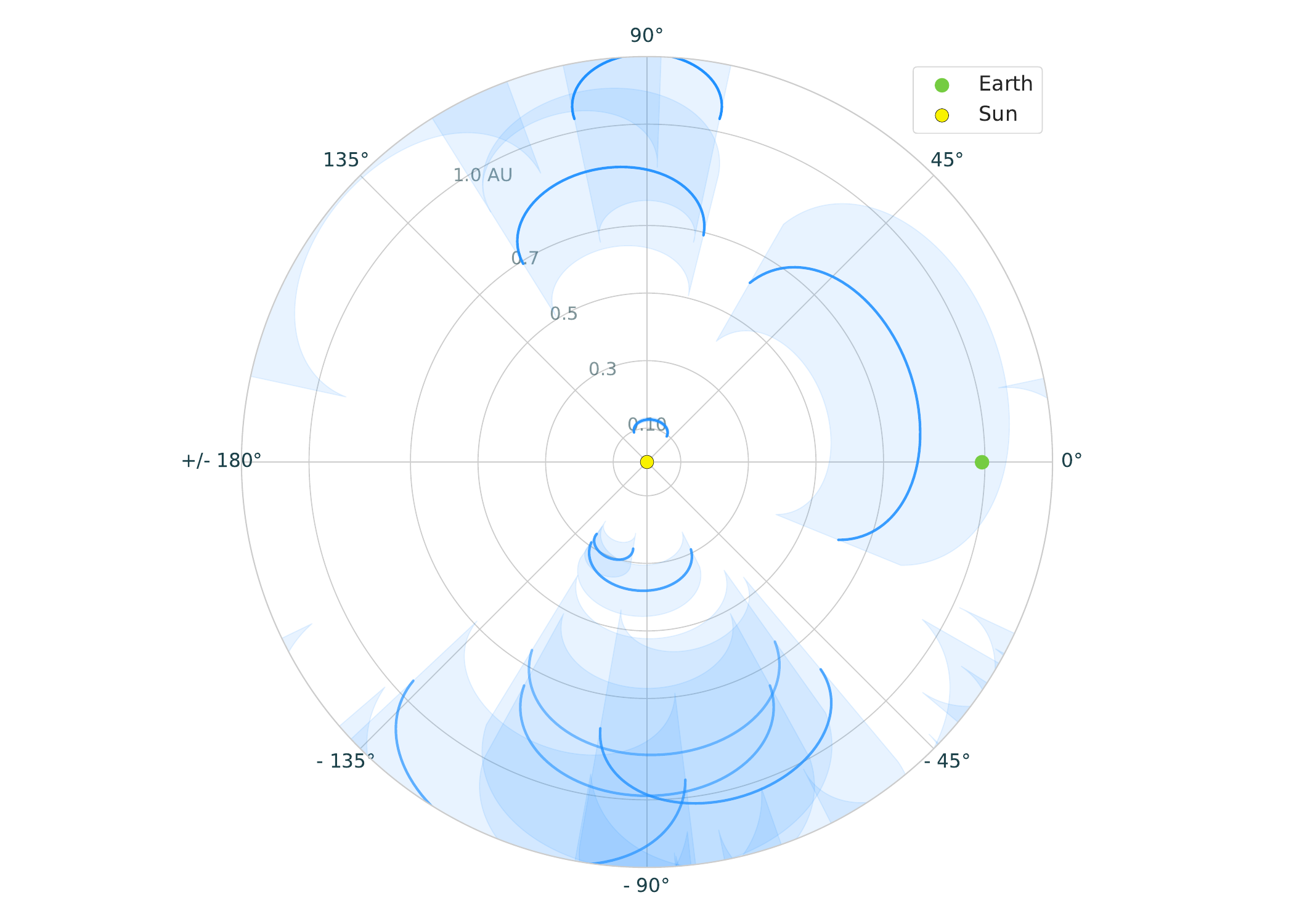}
    \caption{Example visualization of an ELEvo simulation on April 23, 2023 15:00~UT. The yellow dot in the center represents the Sun, and the green dot indicates Earth's position. Each blue semi-ellipse corresponds to a CME, with the solid blue line showing the ensemble mean and the shaded blue area representing $\pm2$ standard deviations across the ensemble members.}
    \label{fig:elevo}
    \end{center}
\end{figure}

An example visualization of an ELEvo multi-CME simulation on April 23, 2023 is shown in Figure~\ref{fig:elevo}, in which several CME fronts are propagated simultaneously, as commonly occurs in operational settings when multiple events are present in the heliosphere. For each event the solid half-ellipse indicates the ensemble mean CME front and the shaded region represents the $\pm 2\sigma$ spread resulting from variations in the input parameters.

\subsubsection{Evaluation of ELEvo}

While ELEvo has been routinely applied in case studies \citep[e.g.][]{weiler_2025_firstobservationsgeomagnetic, davies_2025_realtimepredictiongeomagneticb} and operational settings, it has not yet undergone extensive validation. To address this gap, we evaluate ELEvo using past events listed on the CCMC Scoreboard, a platform which allows space weather offices and scientists to submit their arrival time predictions for CMEs. Specifically, we review each entry in the DONKI catalog to determine whether it is associated with an interplanetary shock (IPS) observed at L1. However, only 196 out of the 3870 events in our cleaned DONKI catalog (see details in Section~\ref{sec:data}) are actually associated with an IPS registered at L1, which is a surprisingly low number. Due to this limitation, we do not assess ELEvo's hit/miss performance. That is, we do not evaluate how many of the predicted arrivals actually occurred, how many false alarms were issued, or how many CMEs that arrived at L1 were missed by ELEvo. Instead, we compare the mean expected arrival time $t_e$ predicted by ELEvo with the actual registered arrival time.

\begin{figure}
    \begin{center}
    \includegraphics[width=\textwidth]{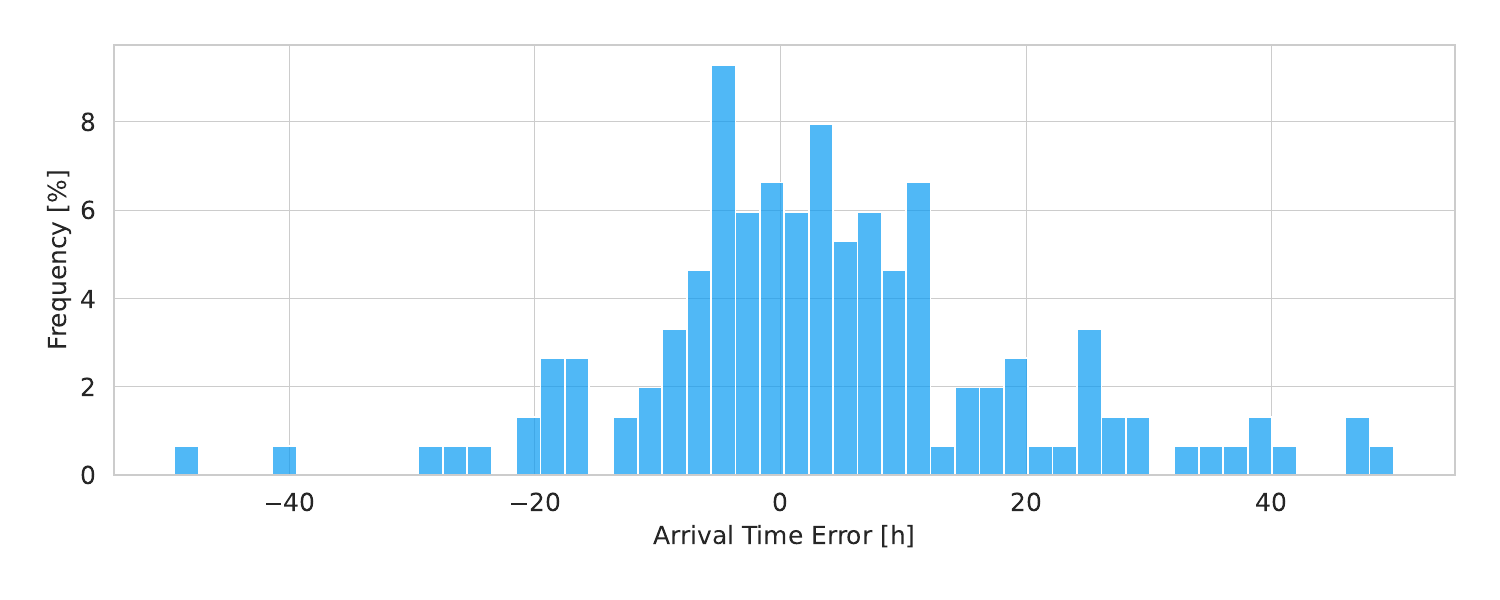}
    \caption{Distribution of arrival time prediction errors from ELEvo simulations, compared against observed IPS arrival times listed in the DONKI catalog.}
    \label{fig:arrival_error}
    \end{center}
\end{figure}

Of the 196 IPSs observed, ELEvo predicted 149 of the associated events to arrive at Earth. The remaining events were not predicted to impact Earth, as their propagation direction and angular half-width placed Earth outside their modeled path under the assumption of self-similar expansion. A comparison between the model-predicted mean arrival times and the actual observed arrival times for these 149 events yields a mean absolute arrival time error of approximately $11.7$ hours, with a standard deviation of around $11.3$ hours, results which are comparable to other arrival time prediction methods currently available \citep[e.g.][]{riley_2018_forecastingarrivaltime, kay_2024_updatingmeasurescme}. The distribution of these arrival time errors is shown in Figure~\ref{fig:arrival_error}.

\subsection{ARCANE}\label{sec:arcane}

The setup of the ARCANE early detection module is described in detail in \cite{rudisser_2026_arcaneearlydetectioninterplanetary}. In brief, the system uses a deep learning model that processes sliding windows of in situ solar wind data and outputs a time series indicating the probability of an ICME being present at each time step.

Each input window spans 1024 time steps of 10 minutes, which corresponds to slightly more than seven days of data. In a real-time operational setting, these windows are advanced step by step as new data becomes available, allowing for continuous classification of the most recent 1024 time steps. Because individual time steps are included in multiple overlapping windows, each one receives multiple classification outputs. To aggregate this information, we compute the average probability across all overlapping outputs for each time step. This approach differs from \cite{rudisser_2026_arcaneearlydetectioninterplanetary}, where each prediction was treated independently to assess the effect of waiting time before classification. In our case, we limit the aggregation to at most 120 outputs per time step, as our experiments show that including additional predictions does not significantly alter the result.

In the present study, we expand upon the original setup of \cite{rudisser_2026_arcaneearlydetectioninterplanetary} by changing the task from binary to multiclass classification, therefore retraining the model from scratch. Previously, the model only distinguished between ``ICME" and ``no ICME". We now separate the CME into sheath and MO, resulting in three possible classes: ``background", ``sheath" and ``MO". To accommodate for this change, the final output layer of the network was modified to produce three class probabilities instead of two, and the loss function was changed from Dice loss to cross-entropy. To assign a definitive label to each time step, we take the class with the highest predicted probability among the three. Other large-scale solar wind structures, such as stream interaction regions, are not explicitly represented as a separate class in the training labels. In the underlying catalog, only CME-related structures are labeled, while all other solar wind structures are formally treated as background. However, since SIRs can exhibit signatures that resemble ICMEs, some ambiguity may arise and misclassifications cannot entirely be excluded. A detailed assessment of such cases is beyond the scope of the present work but has been discussed in related studies \citep[e.g.][]{nguyen_2025_simultaneousmulticlassdetection}.

As outlined in Section~\ref{sec:data}, the evaluation of the complete NEXUS pipeline requires the simultaneous availability of all three data sources (L1 in situ data, an ICME catalog and remote-sensing data). However, training the ARCANE detection module only requires in situ solar wind data and a corresponding ICME catalog, allowing us to use additional years of data for training purposes.

In \cite{rudisser_2026_arcaneearlydetectioninterplanetary}, we employed a nested cross-validation scheme to assess model performance. Here, we adopt a similar approach to make efficient use of the available data. For each target year that we intend to analyze using the NEXUS pipeline, we train a dedicated model that has not been exposed to any data from that year. To avoid potential data leakage at the temporal boundaries, we also exclude the year immediately before and after the target year from training. While this exclusion is conservative, it reflects the fact that solar wind conditions can exhibit persistence on timescales longer than individual events and we aim to ensure that the model cannot overfit to structures or background conditions that later appear in the evaluation period \citep[e.g.][]{owens_2013_27daypersistence}.

The remaining data (i.e., all other years) is divided into three folds for cross-validation: we train three models, each using a different fold for validation and the remaining two for training. The final prediction for the target year is then obtained by averaging the outputs from these three models, resulting in a more robust output.

\subsubsection{Evaluation of ARCANE}\label{sec:arcane_eval}

\begin{figure}
    \begin{center}
    \includegraphics[width=\textwidth]{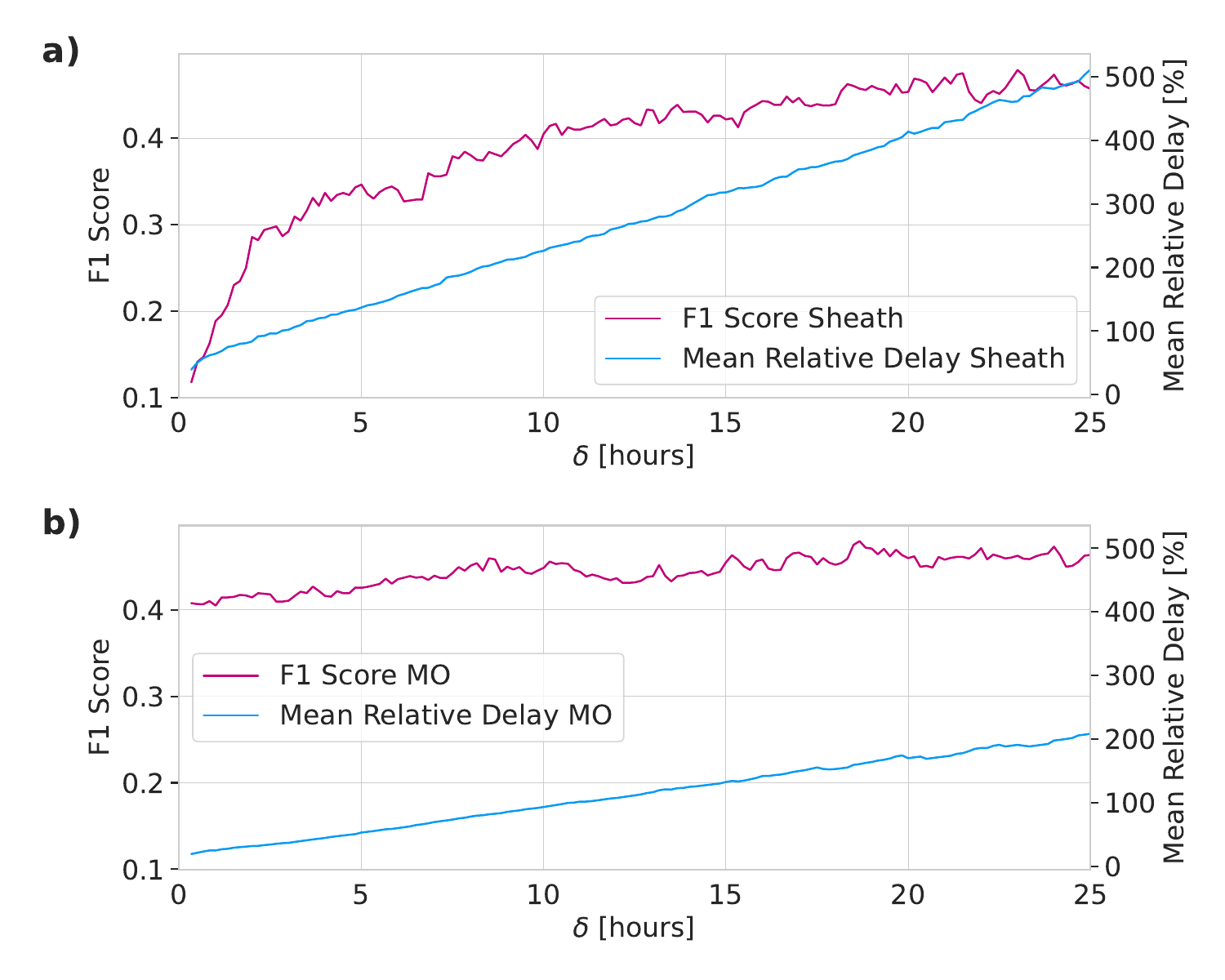}
    \caption{Overview of detection performance as a function of the waiting time $\delta$. (a) F1 Score and mean relative delay (expressed as a percentage of the event duration) as a function of $\delta$ for the sheath region. (b) F1 Score and mean relative delay (in percent of duration) as a function of $\delta$ for the magnetic obstacle (MO).}
    \label{fig:arcane_eval}
    \end{center}
\end{figure}

The early detection capabilities of the ARCANE module were extensively evaluated in \cite{rudisser_2026_arcaneearlydetectioninterplanetary}, where an F1 Score of $0.37$ was achieved at a relative mean delay of $24.1$\% of the event duration. This performance could be further improved to an F1 Score of $0.48$ by awaiting additional data, corresponding to a mean relative delay of $82.9$\% of the event duration.

A similar analysis was conducted here to assess the performance of ARCANE when tasked with detecting sheath and MO as separate substructures of the CME. For both components, we compute the F1 Score as a function of the waiting time $\delta$, which defines how long the model waits before issuing a classification for a given time step after it is first observed. The results are presented in Figure~\ref{fig:arcane_eval}.

The detection performance for both substructures is overall comparable to the results reported in \cite{rudisser_2026_arcaneearlydetectioninterplanetary}. For both sheath and MO detections, the F1 Score increases with increasing $\delta$ and reaches values close to $0.48$ once a substantial fraction of the respective structure has passed (maximum F1 Score of $0.48$ for the sheath at $\delta \approx 23$ h and $0.48$ for the MO at $\delta \approx 18.7$ h). 

For MOs, the F1 Score starts at a higher value for short waiting times and shows smaller relative gains as $\delta$ increases. This suggests that, by the time the model attempts to detect the MO, it has in a lot of cases already observed the sheath and can therefore base the classification on more context. This interpretation is consistent with \cite{rudisser_2026_arcaneearlydetectioninterplanetary}, where we found that the onset of the MO is often required to reliably distinguish an ICME from a driverless shock or other solar wind transients. We note, however, that we did not explicitly quantify whether detection performance differs between sheath-associated and sheathless CMEs in this study. The increased mean relative delay is largely a consequence of splitting CMEs into their substructures. By treating the sheath and MO separately, the effective duration of each target interval is shortened, which inherently increases the relative delay metric. 

A further difference from \cite{rudisser_2026_arcaneearlydetectioninterplanetary} is that, in the present study, we apply a fixed classification threshold of 0.5 for both sheath and MO detections rather than optimizing thresholds. This threshold is therefore not tuned to maximize the F1 Score or to favor either precision or recall. This choice reflects the transition from a binary to a multiclass classification setup, where each time step is assigned to the class with the highest predicted probability. At the maximum F1 Score, for the sheath, a Precision of $0.53$ and a Recall of $0.44$ are achieved. For the MO, the maximum F1 Score corresponds to a Precision of $0.46$ and a Recall of $0.50$.

It is important to note that this version of ARCANE is used in combination with ELEvo, which provides preselected arrival windows. Within this context, a more sensitive detection strategy may be advantageous, as it reduces the risk of missing a relevant event. Furthermore, it should be recognized that ground truth CME catalogs are inherently subjective and reflect the interpretation of individual experts. While reproducing labeled data remains the goal of supervised machine learning, performance must be evaluated in the context of its intended application. In this case, a slightly reduced agreement with a conservative catalog may be acceptable if it supports operational applicability.

\subsection{3DCORE}\label{sec:3DCORE}

In the following, we first describe the general functionality of the 3DCORE model as it has been used in previous studies for retrospective (hindcast) analyses of CME magnetic structure based on complete in situ observations. We then describe how this framework is adapted in the present work for real-time application, where only partial in situ data is available and the goal is short-term forecasting of the remaining magnetic field profile.

The 3DCORE model is a semi-empirical flux rope model initially introduced in \cite{mostl_2018_forwardmodelingcoronal} and subsequently refined in \cite{weiss_2021_analysiscoronalmass, weiss_2021_multipointanalysiscoronal}. 3DCORE was designed to model the characteristic rotation of the magnetic field often observed in CMEs. The model assumes a Gold-Hoyle-type magnetic field configuration embedded within a self-similarly expanding toroidal structure that remains magnetically connected to the Sun during its entire propagation. To simulate propagation effects such as ``pancaking,” 3DCORE allows for elliptical cross-sections. The typical acceleration or deceleration of CMEs due to interaction with the ambient solar wind is modeled via a drag-based approach following \cite{vrsnak_2013_propagationinterplanetarycoronal}. 3DCORE requires 14 input parameters to simulate a CME that fall into five categories:

\begin{enumerate}
    \item \textbf{Launch conditions:} launch time ($t_0$), launch radius ($r_0$), initial velocity ($V_0$)
    \item \textbf{Propagation direction:} longitude ($lon$), latitude ($lat$), inclination ($inc$)
    \item \textbf{Flux rope geometry:} diameter at 1~au ($D_{1\mathrm{au}}$), cross-sectional aspect ratio ($A$, referred to as $\delta$ in \cite{weiss_2021_analysiscoronalmass} and subsequent studies, but renamed here to avoid confusion with the waiting time $\delta$ introduced in Section~\ref{sec:arcane})
    \item \textbf{Expansion and propagation:} expansion rate ($n_a$), drag coefficient ($\Gamma$), background solar wind speed ($v_{\mathrm{sw}}$)
    \item \textbf{Magnetic properties:} twist factor ($T_{\mathrm{f}}$), magnetic field decay rate ($n_b$), field strength at 1~au ($B_{1\mathrm{au}}$)
\end{enumerate}

The twist factor $T_{\mathrm{f}}$ determines the number of magnetic field line rotations along the flux rope and the sign indicates the chirality of the structure: negative values denote left-handed, while positive values correspond to right-handed configurations. The parameters $n_a$ and $n_b$ define how the flux rope diameter and internal magnetic field strength scale with heliocentric distance, following empirical power-law relations from \cite{leitner_2007_consequencesforcefreemodel}.

Beyond its use as a forward model to study CME magnetic structure \citep{rudisser_2024_understandingeffectsspacecraft}, 3DCORE is frequently applied in case studies where it is fitted to in situ observations \citep[e.g.][]{davies_2021_situmultispacecraftremote, telloni_2021_studytwointeracting, mostl_2022_multipointinterplanetarycoronal, long_2023_eruptionmagneticflux, davies_2024_fluxropemodeling, weiler_2025_firstobservationsgeomagnetic}. The fitting is performed using an Approximate Bayesian Computation - Sequential Monte Carlo (ABC-SMC) algorithm.

This algorithm is a particle filtering method that involves generating a large ensemble of input parameter combinations which are used to simulate CMEs via forward modeling. The simulated results are evaluated at specific points (typically spacecraft locations), and their agreement with in situ measurements obtained at this point is quantified using a normalized root mean square error (N-RMSE, $\epsilon$). Input parameter sets that do not produce valid signatures at the defined location are discarded.

The algorithm proceeds iteratively. After each iteration, the parameter space is refined based on the best performing solutions from the previous iteration, thereby gradually converging to a narrowed parameter range. The final result is not a single-best-fit solution, but an ensemble of parameter combinations that approximates the posterior distribution given the observations and their assumed uncertainties. Evaluating these at the point of interest, a mean and associated uncertainty can be derived, typically interpreted as a Normal distribution characterized by its mean and $2\sigma$ spread. Note that the underlying posterior distribution, as characterized by the resulting ensemble of parameter combinations, does not necessarily need to resemble a multivariate Normal distribution so that the derived mean and standard deviation values can in certain scenarios be deceiving. 

To perform a fitting, each model parameter must either be fixed or assigned a range of allowed values to be sampled from during the generation of the ensembles. Broader parameter ranges increase the exploration space and generally result in longer computation times before convergence. To reduce this overhead, parameters that can be constrained using external observations, such as longitude or latitude, which can be estimated from remote-sensing observations near the Sun, are restricted accordingly. Additionally configurable parameters of the fitting algorithm are the number of ensemble members, the minimum and maximum number of iterations and the target N-RMSE. After each iteration, a number of stopping criteria is evaluated:

\begin{itemize}
    \item Has the predefined maximum number of iterations been reached?
    \item Has the minimum number of iterations been reached, and if so, has the target N-RMSE $\epsilon$ been met?
    \item Does the refined parameter space still produce valid model solutions that generate an in situ signature at the spacecraft location within the specified time interval?
\end{itemize}

In addition to the launch time $t_0$, the fitting procedure requires the definition of several points in time. The start time $t_s$ and end time $t_e$ are given to determine the time window in which the CME was observed in situ. Any set of input parameters that produces signatures outside of this window is not considered a valid solution. In addition, a small number (typically fewer than 6) of fitting points ($t_i$) are set, which define the exact times at which the model output is compared to the in situ data during the optimization. While the underlying data have a cadence of 10 minutes, using a much denser set of fitting points does not necessarily improve the reconstruction, as the dominant uncertainties arise from the simplified flux rope geometry and magnetic field assumptions \citep{weiss_2021_multipointanalysiscoronal}. Moreover, the computational cost of the fitting procedure scales approximately linearly with the number of fitting points, since the model–data comparison must be evaluated for each point during every iteration of the optimization. A small number of representative points is therefore sufficient to constrain the large-scale magnetic field rotation while maintaining computational efficiency for the real-time application considered here. These points can be equally spaced, but are often chosen strategically to emphasize specific features of the CME.

\begin{figure}
    \begin{center}
    \includegraphics[width=\textwidth]{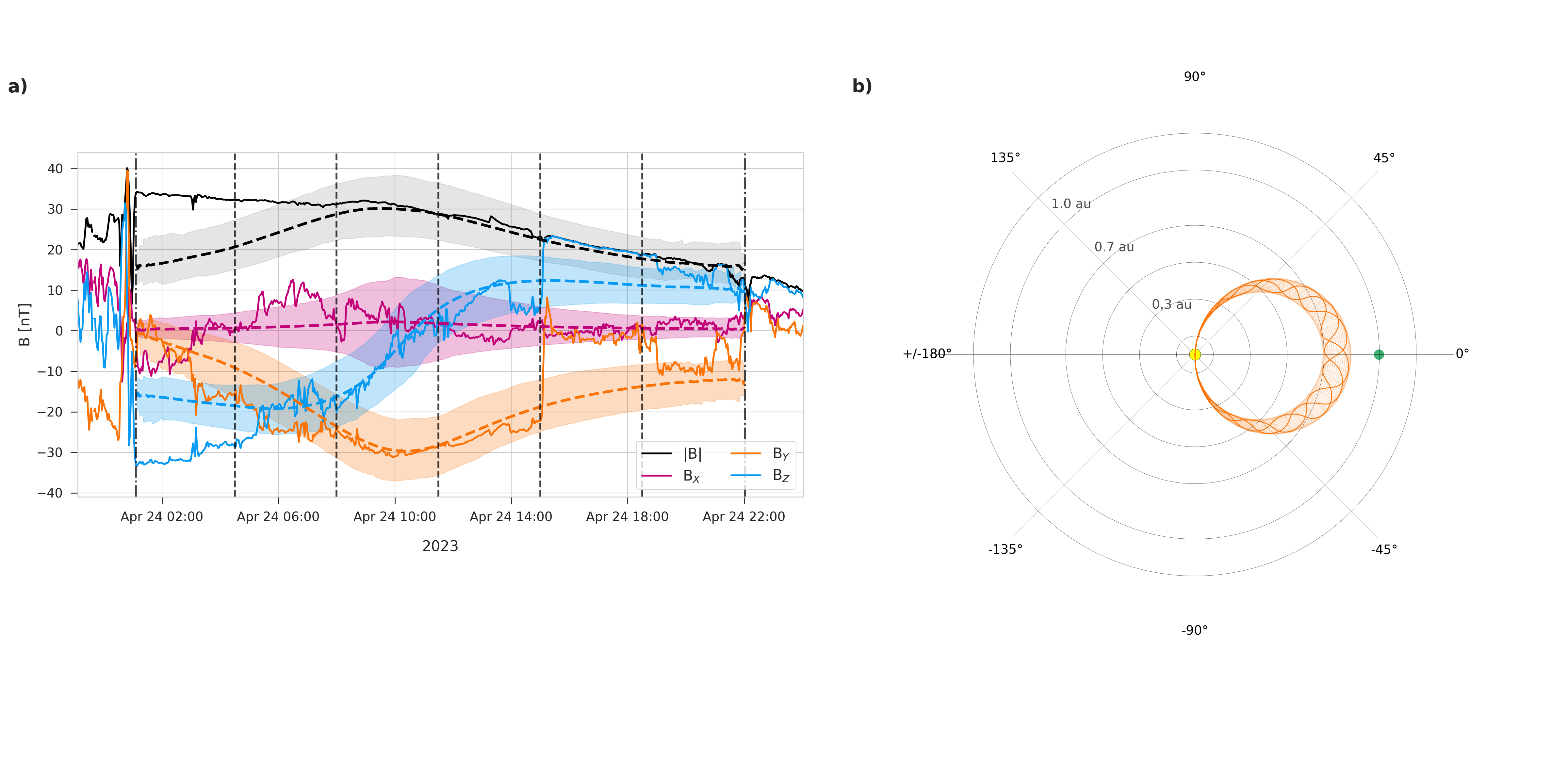}
    \caption{Example 3DCORE reconstruction of the event launched on April 21, 2023, observed by Wind. (a) In situ magnetic field measurements (solid lines) together with the corresponding 3DCORE fitting results. The shaded region indicates the $2 \sigma$ spread of the ensemble, and the dashed colored lines represent the ensemble mean. The vertical black dashed lines mark the start and end times of the MO, which define the reconstruction interval. The vertical gray dashed lines denote the fitting points used to evaluate the quality of the fit throughout the iterative optimization. Magnetic field components are shown in geocentric solar magnetic (GSM) coordinate system. (b) Top down view on the solar equatorial plane showing the 3D mean 3DCORE solution at the time of the observation at Wind (2023 April 23 15:06~UT), 10 hours before the MO arrived at Wind. The orange solid lines within the 3D shape represent three specific field lines of the CME.}
    \label{fig:3dcore_example}
    \end{center}
\end{figure}

An example reconstruction of the event launched on April 21, 2023, reaching 21.5~\rsspace at 21:09~UT, which impacted Wind at a heliocentric distance of 1.0~au, is shown in Figure~\ref{fig:3dcore_example}. The start of the MO was determined to occur on April 24, 2023 01:06~UT and the end time of the MO was set to April 24, 2023 at 22:02~UT, as given in the corresponding ICMECAT entry.

Using 3DCORE in this fitting framework enables the retrospective analysis of individual CME events, in which the model is constrained directly by in situ measurements. In this way, pointwise observations can be exploited to infer the global large-scale structure of the flux rope and its evolution. In cases where several spacecraft observe the same event, the approach can be extended to multipoint analyses \citep[e.g.][]{davies_2021_situmultispacecraftremote, weiss_2021_multipointanalysiscoronal,mostl_2022_multipointinterplanetarycoronal, long_2023_eruptionmagneticflux, davies_2024_fluxropemodeling}, providing a means to assess how coherent CME structures remain over large spatial and temporal scales. Insights gained from such case studies provide valuable constraints for improving CME modeling and enhancing our understanding of the large-scale structure of CMEs.

While such hindsight analyses are highly informative, space weather forecasting requires knowledge of the magnetic structures before the full CME has passed the observer, in order to assess its geoeffectiveness ahead of time. To address this, instead of fitting the model to the complete in situ time series, 3DCORE can be fitted to only an initial portion of the flux rope. The fitted model is then used to predict the yet unseen remainder of the structure. As more data becomes available while the CME sweeps over the observer, the forecast can be iteratively updated. This represents a real-time application of the method rather than a purely hindsight analysis. We refer to this approach as ``Short-Term Forecast", a concept introduced and tested for the first time in this work.

\begin{figure}
    \begin{center}
    \includegraphics[width=\textwidth]{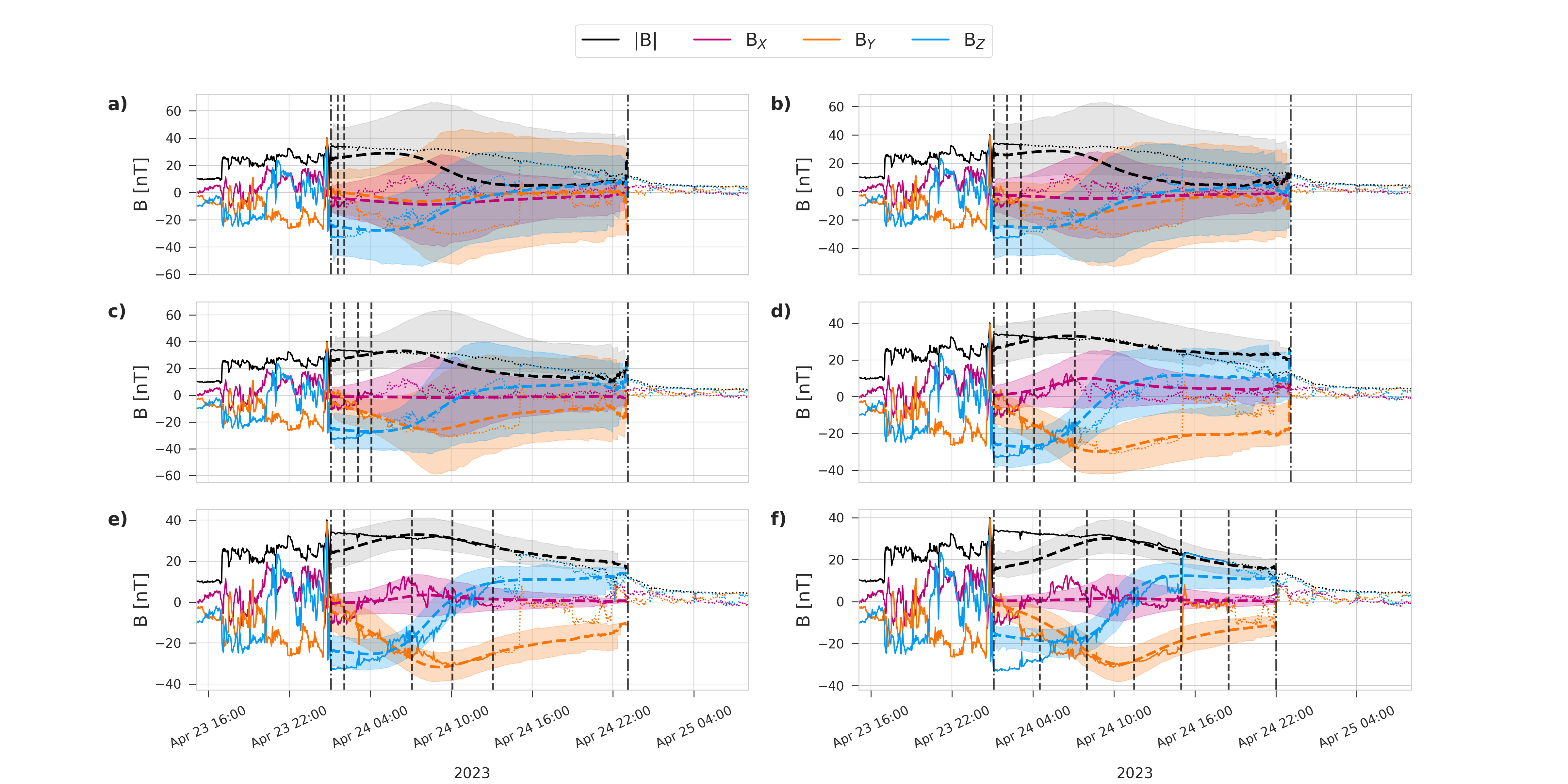}
    \caption{Example showing the iterative 3DCORE short-term forecast procedure for the event launched on April 21, 2023, observed by Wind on 2023 April 24, also used as an example in Figure~\ref{fig:3dcore_example}a. The fitting results are shown after: (a) 1 hour, (b) 2 hours, (c) 3 hours, (d) 6 hours, (e) 12 hours and (f) for the complete event.}
    \label{fig:shortterm_forecast_example}
    \end{center}
\end{figure}

\subsubsection{Evaluation of 3DCORE}

\cite{weiss_2021_analysiscoronalmass} conducted a proof-of-concept study to evaluate the performance of 3DCORE. In this work, synthetic in situ data was generated using 3DCORE, artificial noise was added and the ABC-MC algorithm was used to assess whether the original model input parameters could be reliably recovered. Since the true parameters were known in this setup, the study served as an initial test of the model's potential for parameter estimation. Beyond this, 3DCORE has been applied in a limited number of case studies \citep[e.g.][]{davies_2021_situmultispacecraftremote, telloni_2021_studytwointeracting, mostl_2022_multipointinterplanetarycoronal, long_2023_eruptionmagneticflux, davies_2024_fluxropemodeling, weiler_2025_firstobservationsgeomagnetic}, with varying levels of success depending on the clarity of the observed flux rope signatures. However, due to a lack of definitive ``ground truth" parameters in real cases, a comprehensive validation of the model has not yet been feasible and no statistical assessment of its applicability has been carried out to date.

In many single-spacecraft applications, 3DCORE performs well, particularly when the flux rope signatures are clear and relatively undisturbed. However, its limitations become more apparent in multipoint studies \citep[e.g.][]{palmerio_2021_cmemagneticstructure, weiss_2021_multipointanalysiscoronal,davies_2024_fluxropemodeling} where observations from widely separated spacecraft reveal that such a simple flux rope representation cannot fully capture the complexity of CMEs across large spatial and temporal scales. This is consistent with recent work by \cite{al-haddad_2025_magneticfieldstructure}, indicating that the classical picture of a highly twisted circular cross-section magnetic flux rope model is outdated and a more elaborate and realistic representation that better reflects the true complexity of the magnetic ejecta within CMEs is needed. 

In response to this paradigm shift, more sophisticated models that allow for deformable shapes are being actively developed \citep{weiss_2022_writhedanalyticalmagnetic, weiss_2024_distortedmagneticflux}. These models have great potential to represent CME structure more realistically, but their increased flexibility necessarily introduces a larger number of free parameters. As a consequence, issues of parameter non-uniqueness (i.e., different parameter combinations yielding essentially identical in situ signatures, also known as degeneracy) can arise, especially when only limited observational constraints are available \citep{riley_2004_fittingfluxropes, lynch_2022_utilityfluxrope}.

For these reasons, we deliberately adopt a comparatively simple flux rope model in the present work. Its reduced number of parameters enables efficient fitting in an operational context without relying on heavy regularization or strong prior assumptions. The goal here is therefore not to capture the full complexity of CMEs, but to evaluate how a simple model can be employed for practical short-term forecasting applications. More complex or deformable models could be integrated into this framework in the future, but their benefit will depend on the availability of additional observational constraints capable of resolving meso-scale CME structure.

%%%%%%%%%%%%%%%%%%%%%%%%%%%%%%%%%%%%%%%%%%%%%%%
%  PIPELINE
%%%%%%%%%%%%%%%%%%%%%%%%%%%%%%%%%%%%%%%%%%%%%%%

\section{The NEXUS Pipeline}\label{sec:pipeline}

As outlined in Section~\ref{sec:method}, our NEXUS pipeline integrates three different models. In a real operational setting, NEXUS is triggered by the appearance of a new CME entry in the DONKI catalog. The launch time at $21.5$~\rs, along with the CME's longitude ($lon$), latitude ($lat$), half-width ($\lambda$), and initial speed ($V_0$), are then extracted from the DONKI entry and used to initialize the ELEvo simulations.

If ELEvo predicts an Earth impact, we extract the predicted window of arrival as described in Section~\ref{sec:elevo}. Once this window begins at L1, we use ARCANE to examine the in situ measurements for CME arrival signatures. 
A CME is considered detected if the onset of a sheath is identified within the predicted arrival window. For events lacking a sheath, the detection of the MO onset within the window counts as a detection instead.

If no valid onset is detected, ARCANE is reapplied as time progresses and additional data becomes available. Once an onset has been confirmed, NEXUS advances to the reconstruction stage. If a sheath is present, NEXUS waits for the onset of the associated MO. However, if no MO is detected within 2 hours after the sheath ends, the two are assumed to be unrelated, and the detection process is restarted.

Once an MO has been observed for at least one hour, an initial reconstruction using the 3DCORE model is performed. At this stage, the distinction between sheath and MO provided by ARCANE becomes essential, as the reconstruction is performed only for the MO. This interval contains the organized flux rope (if present) that the model is designed to reproduce, whereas the sheath serves uniquely as a precursor region used to identify the arrival of a CME and is not included in the reconstruction itself.

In the underlying study, we restrict the input parameters for 3DCORE as given in Table~\ref{tab:fitting_params}. These ranges are chosen in agreement with the values used for initializing ELEvo as they were extracted from the associated DONKI entry, insights gained from the analysis of the CMEs at 1~au listed in the ICMECAT catalog, as well as general experience gained from applying 3DCORE. For example, \cite{kay_2024_collectioncollationcomparison} reported typical differences between two independent reconstructions of the same event of about $4.0^\circ$ in latitude and $8.0^\circ$ in longitude. However, the meta-catalog used in their study combines results from multiple reconstruction techniques, many of which rely on the Graduated Cylindrical Shell (GCS) model \citep{thernisien_2006_modelingfluxrope, thernisien_2009_forwardmodelingcoronal, thernisien_2011_implementationgraduatedcylindrical}. The GCS model allows for greater flexibility in the outer shape of CMEs, for instance through parameters such as the angular width between the ``legs". In contrast, 3DCORE employs a more constrained flux rope geometry. Prior work by \cite{rudisser_2024_understandingeffectsspacecraft} has shown that the resulting in situ profiles are highly sensitive to variations in the input latitude and longitude. For this reason, we adopt parameter ranges closer to the maximum median absolute deviation (maxMAD) reported in \cite{kay_2024_collectioncollationcomparison}, namely $21.2^\circ$ for latitude and $50.1^\circ$ for longitude. Accordingly, we allow ranges of $\pm 20^\circ$ in latitude and $\pm 50^\circ$ in longitude, which provides sufficient flexibility. The initial velocity $V_{0}$ is varied by 115~\kms, in accordance with the typical uncertainty in determining the speed as reported by \cite{kay_2024_collectioncollationcomparison}. To obtain the background solar wind speed $v_{\mathrm{sw}}$, we calculate the mean solar wind speed within the previous 48 hours before the start of the sheath (or the start of the MO if no sheath is associated with it) and vary it by 100~\kms.

\begin{table}
\centering
\caption{Model parameter ranges: longitude ($lon$), latitude ($lat$), inclination ($inc$), diameter at 1 au ($D_{1\mathrm{au}}$), aspect ratio ($A$), launch radius ($r_{0}$), initial velocity ($V_{0}$), expansion rate ($n_{a}$), background drag ($\Gamma$), solar wind speed (\vsw), twist factor ($T_{f}$), magnetic decay rate ($n_b$) and magnetic field strength at 1 au ($B_{1\mathrm{au}}$). The longitude and latitude of the flux rope apex are given in  Heliocentric Earth Equatorial (HEEQ) coordinates.}

\begin{tabular}{l c c c}
    Parameter & Unit  & Minimum  & Maximum  \\[0.5ex]
    \hline
    $lon$ & deg & \donkilon $ - 50$ & \donkilon  $ + 50$  \\[0.5ex]
    $lat$ & deg & \donkilat $ - 20$ & \donkilat $ + 20$\\[0.5ex]
    $inc$ & deg & $0.$ & $360.$ \\[0.5ex]
    $D_{1\mathrm{au}}$ & au & $0.15$ & $0.35$ \\[0.5ex]
    $A$ &  &$2.$ & $2.$ \\[0.5ex]
    $r_{0}$ & \rs &$21.5$ & $21.5$ \\[0.5ex]
    $V_{0}$ & \kms & $V_{0,\mathrm{DONKI}} - 115$ & $V_{0,\mathrm{DONKI}} + 115$ \\[0.5ex]
    $n_{a}$ & &  $1.14$ & $1.14$ \\[0.5ex]
    $\Gamma$ & & $0.1$ & $2.4$ \\[0.5ex]
    $V_{\mathrm{sw}}$ & \kms & $\langle V_{\mathrm{sw}}\rangle_{48\,\mathrm{h}} - 100$ & $\langle V_{\mathrm{sw}}\rangle_{48\,\mathrm{h}} + 100$ \\[0.5ex]
    $T_{f}$ & & $-100.$ & $100.$ \\[0.5ex]
    $n_{b}$ & & $1.64$ & $1.64$ \\[0.5ex]
    $B_{1\mathrm{au}}$ & nT & $5.$ & $75.$ \\[0.5ex]
    \hline
\end{tabular}
\label{tab:fitting_params}
\end{table}

This first reconstruction is repeated at fixed intervals after MO onset (2, 3, 6 and 12 hours), provided that ARCANE has not yet indicated the end of the event. Finally, once ARCANE determines that the MO has ended, a final reconstruction is carried out using the actual start and end times of the event as identified by ARCANE, provided that the event duration exceeds 6 hours. The corresponding fitting points for each reconstruction are listed in Table~\ref{tab:fittingpoints}. It is important to note that, in a real-time scenario, the actual duration of the MO is unknown until the full structure has passed. Therefore, we define the endpoint $t_e$ using a fixed estimated duration $d_e = 22.0$ hours. This value is motivated by the mean MO duration at 1~au, derived from the HELIO4CAST ICME catalog, which is $21.6$ hours. It is important to note that this final reconstruction is not used for forecasting. Instead, it serves as a reference solution to evaluate the performance of the earlier short-term forecasts. In this case, the reconstruction is based on the complete in situ data and uses the event duration determined by ARCANE rather than the estimated duration.

\begin{table}
\centering
\caption{Fitting points $t_i$ for each observed duration $d_{obs}$. $t_s$ denotes the in situ start time of the event, $d$ the actual duration of the event and $d_e$ the estimated duration of the event, which is set to 22 hours.}

\begin{tabular}{l c c c c c c}
    $d_{obs}$ & $t_1$  & $t_2$  & $t_3$  & $t_4$ & $t_5$ & $t_e$ \\[0.5ex]
    \hline
    1 h & $t_s + 30$ min & $t[-1]$ & - & - & - & $t_s+d_e$ h \\[0.5ex]
    2 h & $t_s + 1$ h & $t[-1]$ & - & - & - & $t_s+d_e$ h \\[0.5ex]
    3 h & $t_s + 1$ h & $t_s + 2$ h & $t[-1]$ & - & - & $t_s + d_e$ h \\[0.5ex]
    6 h & $t_s + 1$ h & $t_s + 3$ h & $t[-1]$ & - & - & $t_s + d_e$ h \\[0.5ex]
    12 h & $t_s + 3$ h & $t_s + 6$ h & $t_s + 9$ h & $t[-1]$ & - & $t_s + d_e$ h \\[0.5ex]
    $d$ h & $t_s + \frac{d \times 1}{6}$ h & $t_s + \frac{d \times 2}{6}$ h & $t_s + \frac{d \times 3}{6}$ h & $t_s + \frac{d \times 4}{6}$ h  & $t_s + \frac{d \times 5}{6}$ h & $t_s + d$ h  \\[0.5ex]
\hline

\end{tabular}
\label{tab:fittingpoints}
\end{table}

It is important to note that the event start time may be updated during the observation period, as ARCANE refines its detection based on incoming data. As a result, it is possible that no valid event is detected at some of these reconstruction intervals. An event is assigned to a reconstruction interval if its predicted duration falls within specific bounds. For example, a duration $l$ is assigned to the 1-hour reconstruction if $1 \leq l < 2$, or to the 6-hour reconstruction if $6 \leq l < 12$. A reconstruction is performed as soon as an event of the corresponding duration is detected and is not repeated for that interval at any point.

As described in Section~\ref{sec:3DCORE}, the fitting procedure involves multiple iterations, during which the parameter space is progressively constrained. After each iteration, the NEXUS pipeline checks whether one of the stopping criteria has been met. In this study, the fitting is terminated when one of the following conditions is satisfied: 

\begin{itemize}
    \item a maximum of 15 iterations is reached
    \item a minimum of 12 iterations is completed and the normalized root-mean-square error (N-RMSE) falls below 0.25
    \item a time limit of 20 minutes is exceeded
\end{itemize}

The time constraint is imposed to ensure that the NEXUS pipeline remains suitable for real-time application. If exceeded, it also interrupts the current reconstruction iteration, even if that iteration has not yet been completed.

Initializing 3DCORE with automatically defined parameter ranges and fitting points may, in some cases, result in reduced quality fits compared to manually configuring 3DCORE through trial and error. As discussed earlier, strategically selecting fitting points allows for emphasizing specific signatures that should be represented in the reconstruction, or avoiding outliers in cases of noisy or disturbed data, for example, due to propagation effects or interactions the CME has already undergone. In addition, the twist parameter determining the handedness of the flux rope can be constrained to either positive or negative values by an experienced observer who has already visually identified the handedness \citep{bothmer_1998_structureoriginmagnetic}. Although such strategies are commonly applied in single-event studies, NEXUS is intentionally designed to operate without human intervention or manual adjustments.

\begin{figure}
    \begin{center}
    \includegraphics[width=\textwidth]{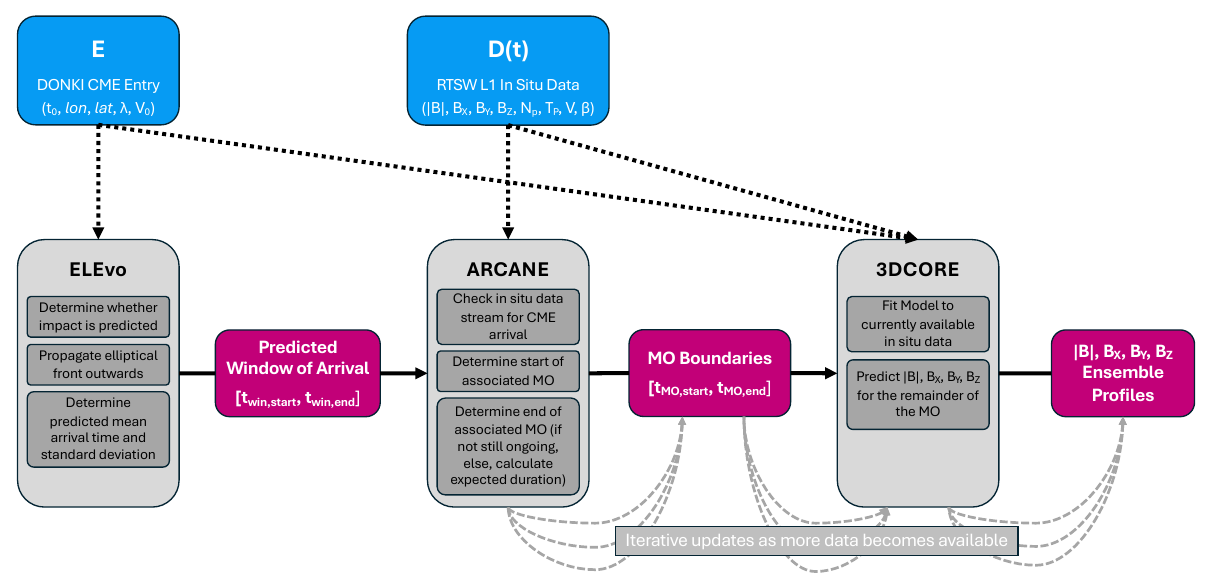}
    \caption{Schematic overview of the operational NEXUS pipeline. Blue boxes denote real-time data products, where $E$ and $D(t)$ correspond to the descriptions in the pseudo code in Algorithm~\ref{alg:pipeline}. Gray boxes represent the three coupled models and their internal components, and magenta boxes indicate model outputs. Dotted lines illustrate which data products are used as inputs by each model, solid lines show how model outputs serve as input for subsequent stages of the pipeline, and dashed gray lines mark steps that are iteratively repeated as additional in situ measurements become available.}
    \label{fig:diagrams}
    \end{center}
\end{figure}

A simplified overview of the NEXUS pipeline, presented in pseudo-code, is provided in Algorithm~\ref{alg:pipeline}. Figure~\ref{fig:diagrams} provides a schematic overview of the corresponding data flow, model coupling and iterative update logic of NEXUS.

\begin{algorithm}
\caption{NEXUS: Operational CME detection and reconstruction pipeline.}
\label{alg:pipeline}

\begin{algorithmic}[1]

\Require New DONKI CME entry $E$; continuous L1 in situ data stream $D(t)$
\Ensure Set of 3DCORE reconstructions $\{R_k\}$

\State \textbf{Detection stage:}
\State Extract CME parameters from $E$:
\Statex \hspace{\algorithmicindent} $t_0$ (21.5 \rs), longitude ($lon$), latitude ($lon$), half-width ($\lambda$), initial speed ($V_0$)

\State Run ELEvo with ($t_0$, $lon$, $lat$, $\lambda$, $V_0$)

\If{ELEvo predicts no Earth impact}
    \State \Return \Comment{event discarded}
\EndIf

\State Obtain predicted arrival window $[t_{\mathrm{win,start}}, t_{\mathrm{win,end}}]$
\State detectionConfirmed $\gets \mathrm{FALSE}$

\While{detectionConfirmed $= \mathrm{FALSE}$}
    \State Wait until new L1 in situ data are available up to current time $t_{\mathrm{cur}}$
    \State Apply ARCANE to $D(t)$ over $[t_{\mathrm{win,start}}, t_{\mathrm{cur}}]$

    \If{sheath onset $t_{\mathrm{sh}} \in [t_{\mathrm{win,start}}, t_{\mathrm{win,end}}]$}
        \State detectionConfirmed $\gets \mathrm{TRUE}$

        \Comment{search for associated MO}
        \State Check if time since sheath end $t_{\mathrm{sh,end}}$ is less than max gap $\Delta t_{\mathrm{sheath\text{-}MO,max}}$ between $t_{\mathrm{sh,end}}$ and MO onset $t_{\mathrm{MO,start}}$
        \While{$t_{\mathrm{cur}} \le t_{\mathrm{sh,end}} + \Delta t_{\mathrm{sheath\text{-}MO,max}}$}
            \State update ARCANE output with new data
            \If{$t_{\mathrm{MO,start}}$ detected}
                \State \textbf{break} \Comment{MO found}
            \EndIf
        \EndWhile

        \If{MO onset $t_{\mathrm{MO,start}}$ not detected}
            \Comment{sheath assumed unrelated -- restart detection}
            \State $t_{\mathrm{win,start}} \gets t_{\mathrm{sh,end}}$
            \State \textbf{continue}
        \EndIf

    \ElsIf{MO onset $t_{\mathrm{MO,start}} \in [t_{\mathrm{win,start}}, t_{\mathrm{win,end}}]$}
        \Comment{CME without detected sheath}
        \State detectionConfirmed $\gets \mathrm{TRUE}$

    \Else
        \Comment{no onset detected yet}
        \If{$t_{\mathrm{cur}} > t_{\mathrm{win,end}}$}
            \State \Return \Comment{no CME detected within predicted arrival window, event discarded}
        \Else
            \State \textbf{continue}
        \EndIf
    \EndIf
\EndWhile

\Statex
\State \textbf{Reconstruction stage:}

\State Wait until $t_{\mathrm{cur}} \ge t_{\mathrm{MO,start}} + 1~\mathrm{h}$
\State Run 3DCORE with:
\Statex \hspace{\algorithmicindent} DONKI/ELEvo parameters and $D(t)$ over $[t_{\mathrm{MO,start}}, t_{\mathrm{cur}}]$
\State Store reconstruction $R_1$

\For{$\tau \in \{2, 3, 6, 12\}$ hours}
    \State Wait until $t_{\mathrm{cur}} \ge t_{\mathrm{MO,start}} + \tau$ \textbf{ or } ARCANE signals MO end
    \If{MO not yet ended}
        \State Run updated 3DCORE reconstruction using data up to $t_{\mathrm{cur}}$
        \State Store reconstruction $R_{\tau}$
    \Else
        \State \textbf{break}
    \EndIf
\EndFor

\If{$t_{\mathrm{MO,end}} - t_{\mathrm{MO,start}} \ge 6~\mathrm{h}$}
    \State Run final 3DCORE reconstruction on $[t_{\mathrm{MO,start}}, t_{\mathrm{MO,end}}]$
    \State Store $R_{\mathrm{final}}$
\EndIf

\State \Return $\{R_k\}$

\end{algorithmic}
\end{algorithm}

\begin{figure}
    \begin{center}
    \includegraphics[width=\textwidth]{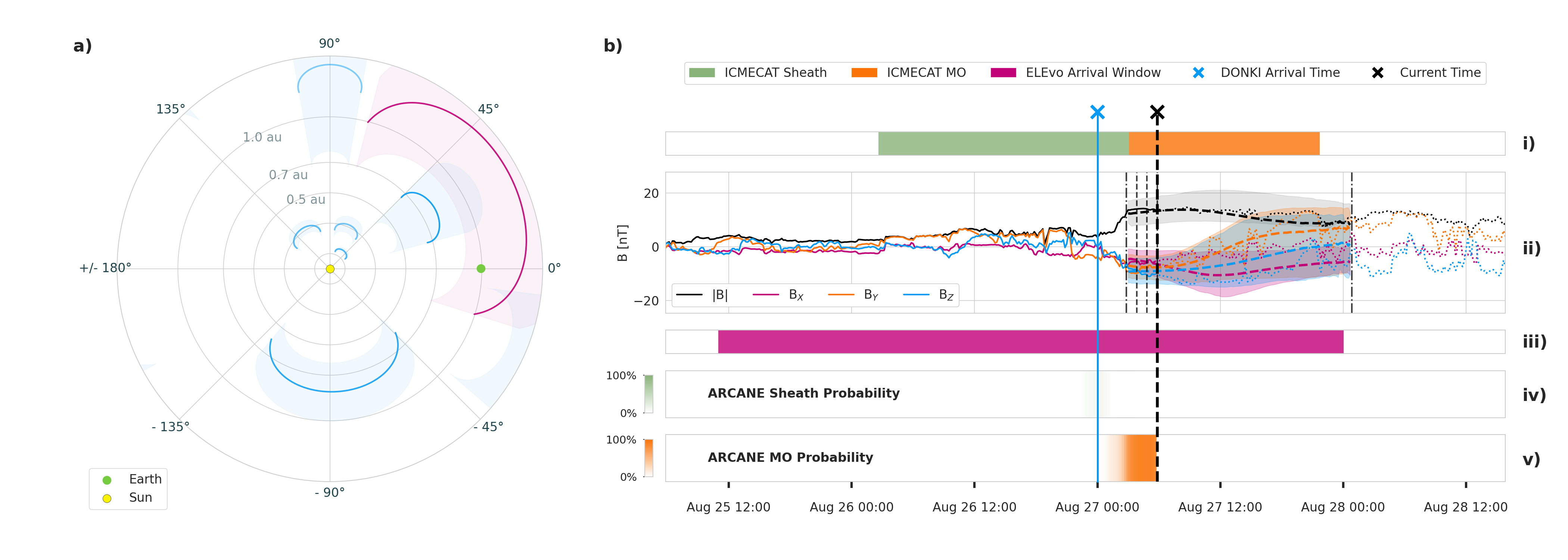}
    \caption{Example of the real-time NEXUS pipeline for the event launched on August 22, 2014 plotted on August 27, 2014 at 05:50~UT, 3 hours after the detected start of the detected MO. Panel (a) shows the ELEvo ensemble simulation with the target CME in magenta and other concurrent CMEs in blue; the shaded region indicates the ensemble spread. Earth (green dot) lies in the trailing part of the ensemble, implying that the predicted arrival window is already in progress. Panel (b) shows the pipeline outputs at L1: (i) the ICMECAT event interval (sheath in green, MO in orange), (ii) RTSW in situ magnetic field data in geocentric solar magnetic (GSM) components, with observations available at the current time shown as solid lines and future data as dotted lines; the ARCANE-detected MO start is marked by the first vertical dash-dotted line, and the MO end, assuming a duration of 22 hours, by the last vertical dash-dotted line. Vertical dashed lines between these boundaries indicate fitting points used for 3DCORE, whose mean solution is shown by dashed curves with the $2\sigma$ spread shaded. Panels (iii)-(v) show, respectively, the ELEvo arrival window (magenta), ARCANE-predicted sheath probabilities (green) and ARCANE-predicted MO probabilities (orange). The blue vertical dashed line with cross indicates the DONKI-reported IPS arrival time, while the black dashed vertical line with cross denotes the current time. At the time shown in the static frame, the ARCANE sheath probability is close to zero over the displayed time interval and therefore appears barely visible in the plot.
    An animated 45~seconds version of this figure is available in the online journal. The animation spans the period from August 23, 2014 at 23:30~UT to August 29, 2014 at 10:00~UT, showing the continuous evolution of the ELEvo simulation together with the available L1 in situ data at the same time. As more data becomes available, the 3DCORE reconstructions and ARCANE probability estimates are correspondingly updated. The static figure corresponds to one frame of this animation.}
    \label{fig:pipeline_example_event}
    \end{center}
\end{figure}

An example event listed in the DONKI database, reaching 21.5~\rs on August 22, 2014 at 19:30~UT, together with the corresponding output of our NEXUS pipeline at August 27, 2014 at 05:50~UT, 3 hours after the detected start of the MO, is shown in Figure~\ref{fig:pipeline_example_event}, with the current time indicated by the vertical black dashed line with the black cross symbol on top in panel~(b). Panel~(a) displays the ELEvo simulation output. The CME associated with the considered event is highlighted in magenta, while other simultaneously active CMEs are shown in blue. The shaded region around the mean trajectory of each event represents the $2\sigma$ ensemble spread. In this example, the ensemble mean arrival time has already elapsed, such that Earth (green dot) lies in the trailing part of the ensemble distribution. Consequently, the ELEvo-predicted arrival window is already partially completed.

Panel~(b) summarizes the associated in situ data and model information at L1. In panels~(i)-(iii) we show contextual timing information. Panel (i) shows the timing for the sheath and MO of the event as given in the ICMECAT, which we treat as ground truth. Panel (ii) shows the RTSW magnetic field in situ data obtained until the current point in time as colored solid lines. The colored dotted lines that continue until the end of the shown time period indicate future in situ data that have not yet been included in the analysis. Also shown is the ARCANE-detected MO start time, obtained after thresholding the ARCANE probabilities, which is indicated by a black vertical dash dot line. The corresponding event end time is estimated from the assumed duration. The corresponding fitting points are shown as dashed black vertical lines in between the start and end time. The resulting 3DCORE ensemble fit and prediction for the remainder of the MO is overplotted on the in situ data. The ensemble mean is shown as dashed lines and the $2\sigma$ spread as shaded areas. Panel~(iii) shows the ELEvo-predicted arrival window (magenta).

Panels (iv) and (v) present the ARCANE probabilities for sheath (green) and MO (orange), respectively, showing that in this case no sheath was detected and the MO has been observed for approximately 3 hours at the time evaluated here. The IPS linked to this event in the DONKI database is indicated by the blue vertical line with the blue cross symbol. In this example, the ICMECAT event places the shock arrival, i.e. the start of the sheath, well before the DONKI-reported arrival time, resulting in an unusually long sheath duration of almost 24 hours. While there is a minimal but hardly visible rise in sheath probability around the DONKI-reported IPS arrival time, the threshold to count this as a sheath has not been exceeded. Immediately afterwards, the MO probability increases. The ARCANE-derived MO onset time therefore lies very close to the MO start time listed in the ICMECAT. The assumed duration of the MO, however, places the trailing edge of the MO slightly later than that reported in the ICMECAT. 

With regard to the quality of the fit and the resulting short-term forecast, the visual agreement between the reconstructed and observed magnetic field components is generally good, with the $B_Z$ component showing the strongest deviation compared with the observed data. The reconstruction predicts an almost immediate recovery toward less negative values after the final fitting point, whereas the observations show a continued decrease over several hours. This results in a longer interval of southward $B_Z$ in the measurements than in the 3DCORE prediction. Therefore, in terms of space weather forecasting, the geomagnetic effect would probably have been underestimated from the modeled magnetic field structure. However, it should be noted that the measured $B_Z$ remains within the predicted 3DCORE ensemble spread throughout the event.

%%%%%%%%%%%%%%%%%%%%%%%%%%%%%%%%%%%%%%%%%%%%%%%
%  RESULTS
%%%%%%%%%%%%%%%%%%%%%%%%%%%%%%%%%%%%%%%%%%%%%%%

\section{Results}\label{sec:results}

\subsection{Event \& Arrival Numbers}

Our complete NEXUS pipeline is executed for a total of $3870$ events between the first entry in the DONKI catalog in October 2013 and the last event listed in the ICMECAT in July 2025. Based on their DONKI-derived parameters, $406$ of these events are predicted by ELEvo to impact Earth. From this subset, $28$ events are excluded due to insufficient in situ data coverage within the predicted window of arrival, resulting in a final sample of $378$ ELEvo-predicted arrival windows used in this study. 

\begin{figure}
    \begin{center}
    \includegraphics[width=\textwidth]{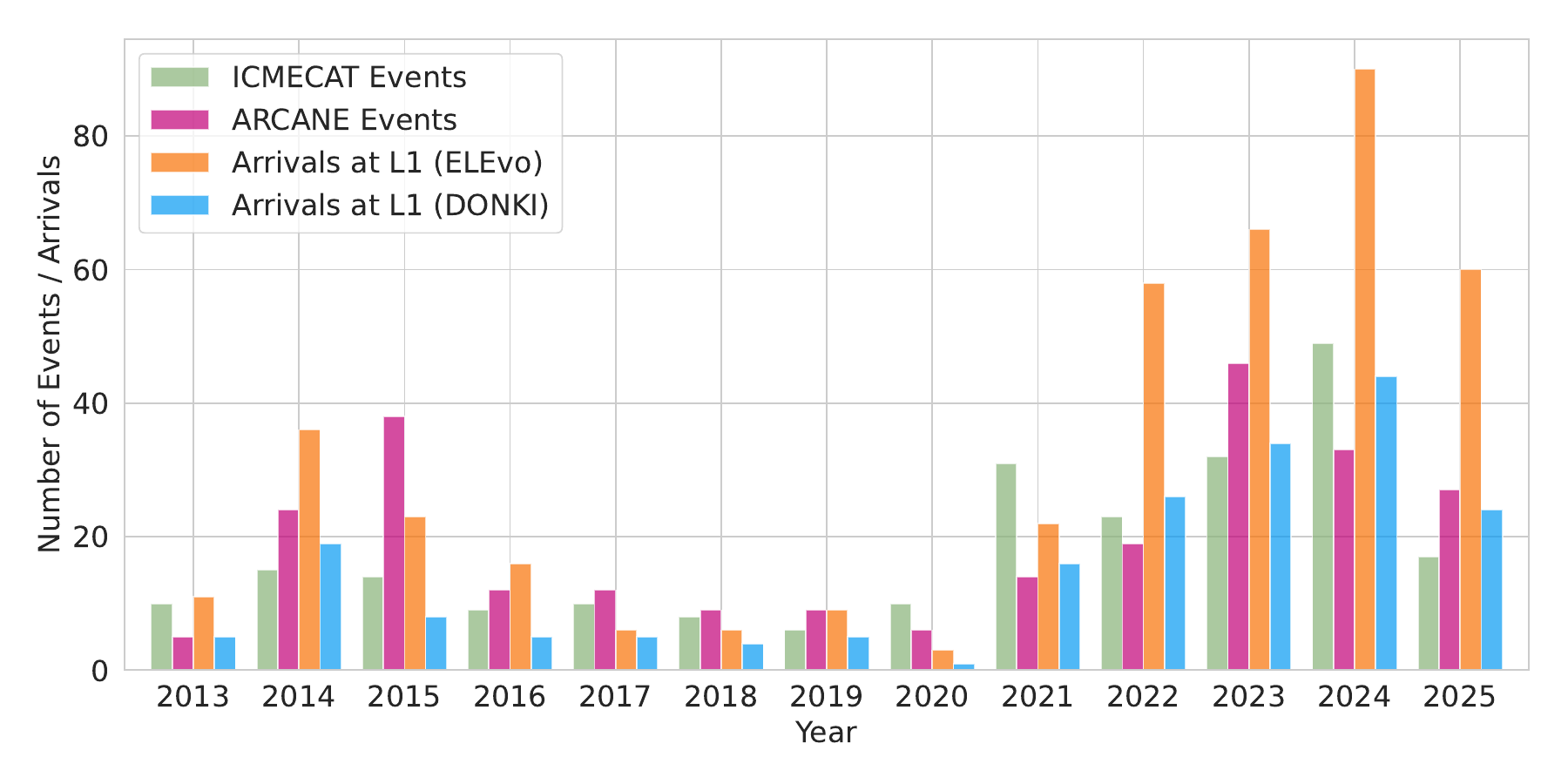}
    \caption{Annual numbers of events and arrivals according to different data sources between 2013 and 2025.}
    \label{fig:event_numbers}
    \end{center}
\end{figure}

For context, and independent of the ELEvo filtering, the ICMECAT lists $234$ ICMEs over the same time period, while DONKI reports $196$ IPS arrivals at L1. ARCANE detects 254 events when applied continuously, independent of whether these detections fall within the ELEvo-predicted arrival time windows. The yearly distribution of all event and arrival counts is shown in Figure~\ref{fig:event_numbers}. Subsequent reconstructions are only performed for events that satisfy both the ELEvo arrival time criterion and the in situ detection, which highlights that the number of events differs substantially among data sources. This reflects the well-known inherent subjectivity in defining ICMEs, which has repeatedly been reported in numerous studies \citep[e.g.][]{richardson_2014_identificationinterplanetarycoronal, rudisser_2026_arcaneearlydetectioninterplanetary, kay_2026_collectioncollationcomparison}. 

\begin{figure}
    \begin{center}
    \includegraphics[width=\textwidth]{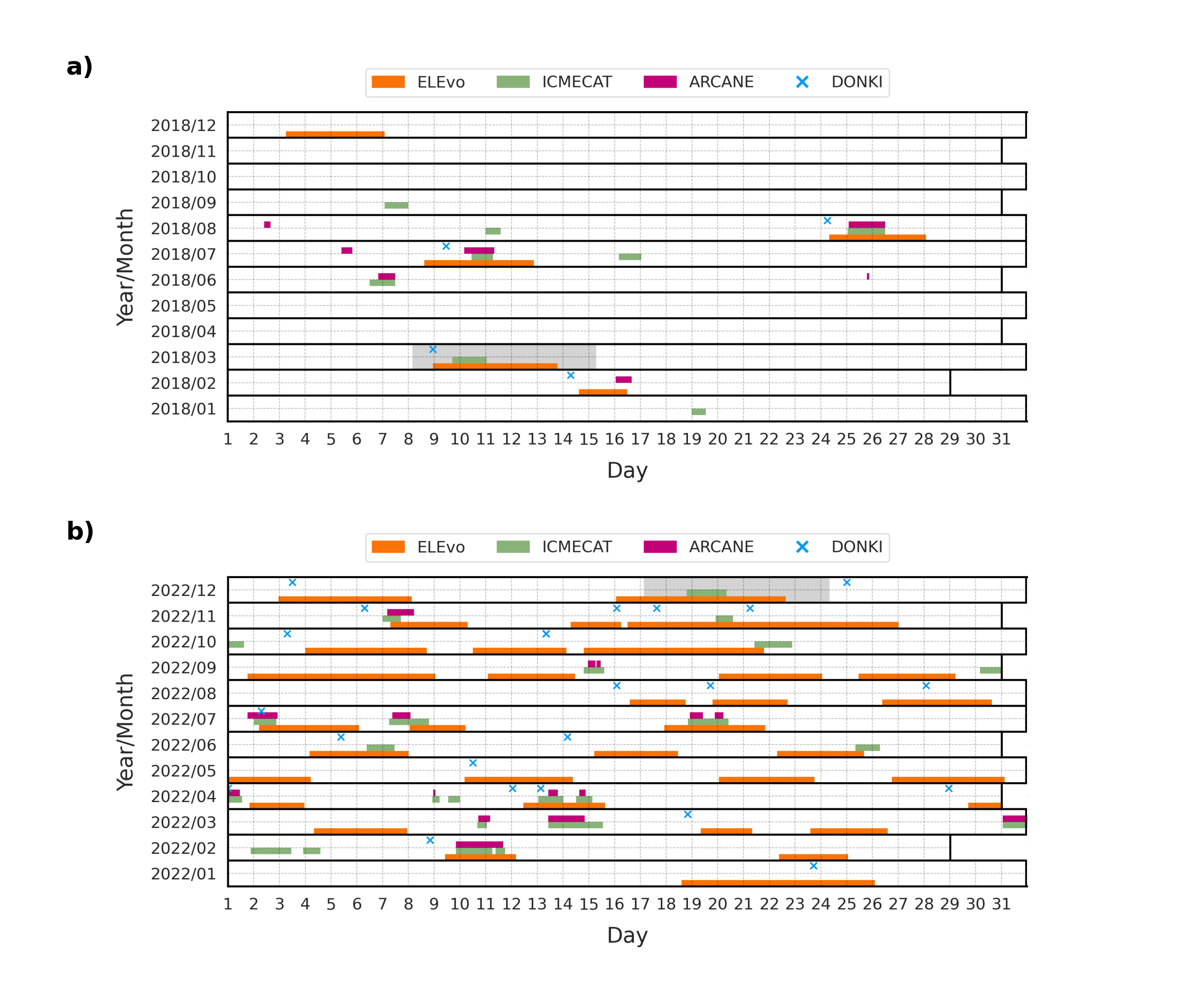}
    \caption{Timeline comparison of events and arrivals for two example years, (a) 2018, (b) 2022. Shown are ELEvo-predicted arrival windows (orange), events (including sheath and MO) listed in the ICMECAT (green), ARCANE-detected events (including sheath and MO, magenta) and DONKI-listed IPS arrivals at L1 (blue crosses). Gray areas indicate intervals without sufficient in situ data coverage.}
    \label{fig:event_numbers_heatmap}
    \end{center}
\end{figure}

To illustrate these discrepancies in more detail, two example years are shown in Figure~\ref{fig:event_numbers_heatmap}. For 2018 and 2022, the figure displays ELEvo-predicted arrival windows, ICMECAT events, ARCANE-detected events and DONKI-listed arrivals. These examples make clear that the correspondence between event lists is often weak. ARCANE and the ICMECAT frequently agree, which is expected given that ARCANE was trained using the ICMECAT as ground truth and therefore aims to reproduce its labeling conventions. However, DONKI arrival times are commonly earlier than the ICME intervals identified in the two other datasets, underlining the ambiguity in identifying shock arrivals. In addition, there are numerous instances in which events appear in only one data source, with no counterpart elsewhere.

\begin{figure}
    \begin{center}
    \includegraphics[width=\textwidth]{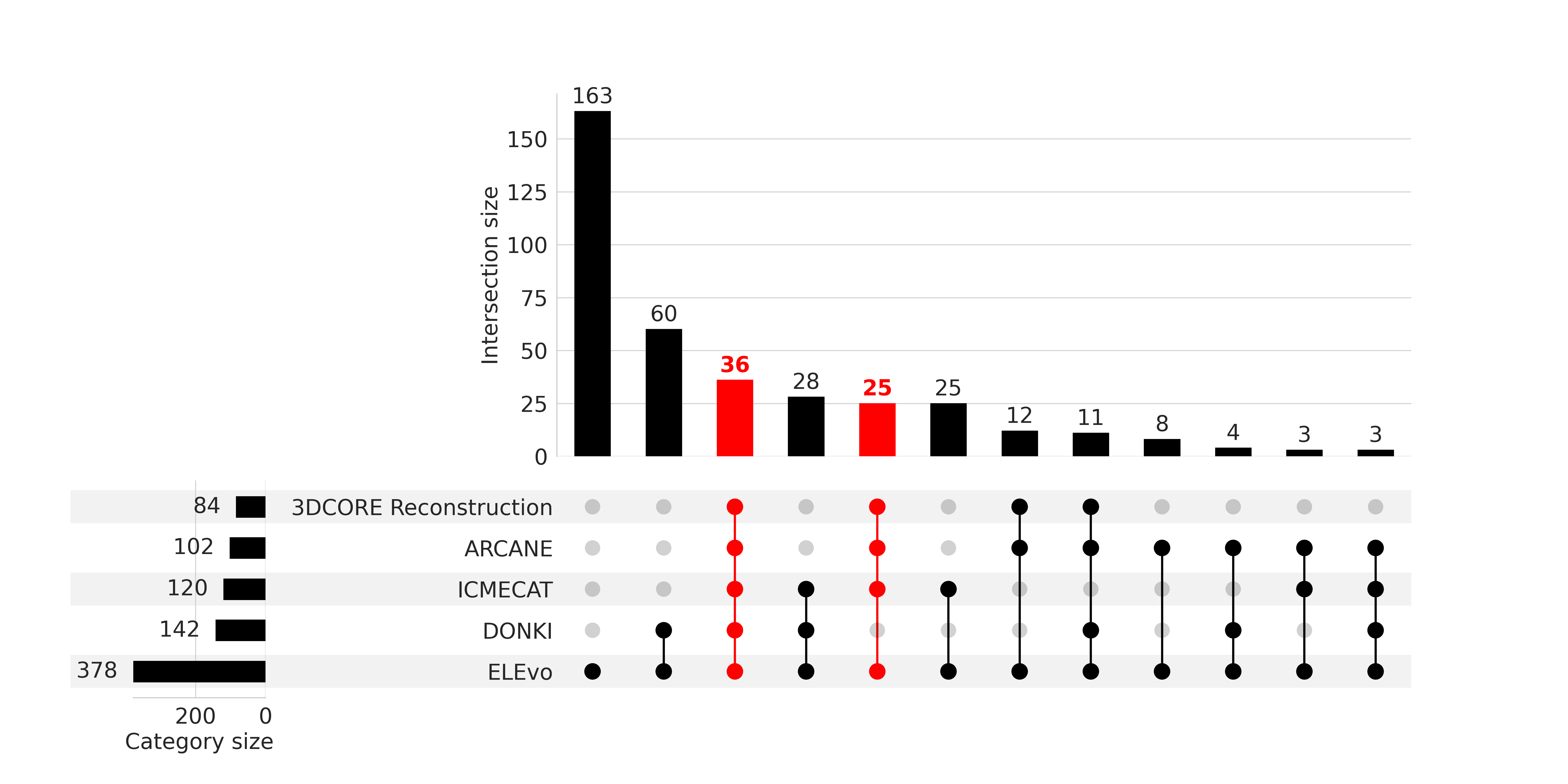}
    \caption{Upset plot illustrating the overlap between DONKI-listed shock arrivals, ICMECAT events, ARCANE detections and their associated reconstructions within ELEvo-predicted arrival windows. Bars show the number of events in each unique combination of data sources, while the connected dots below indicate the contributing sets. The two combinations highlighted in red correspond to the subsets that together form the 61 events used for the final reconstruction quality evaluation, requiring the presence of both an ICMECAT event and a corresponding 3DCORE Reconstruction.}
    \label{fig:upset_plot}
    \end{center}
\end{figure}

The agreement and disagreement patterns between the different data sources are summarized quantitatively in the upset plot shown in Figure~\ref{fig:upset_plot}. The plot is constructed using the set of ELEvo-predicted arrival windows as the reference, and indicates for each window whether it is associated with (i) a DONKI-listed IPS, (ii) an ICMECAT event, (iii) an ARCANE detection, and (iv) a valid 3DCORE reconstruction. An ARCANE-detected event is considered as having a valid reconstruction if at least one of the 6h, 12h, or full-event reconstructions exists. The upset plot emphasizes that only a fraction of ELEvo-predicted arrival windows are consistently associated across all data sources, while a substantial number show partial or no overlap. For example, 163 ELEvo-predicted arrivals have neither a DONKI-listed IPS, ICMECAT event, ARCANE detection or valid 3DCORE reconstruction within the determined window of arrival time and 60 only feature an associated DONKI-listed arrival, without any of the other categories matching the window. Highlighted in red are the two categories that have an ICMECAT event, an ARCANE detection and a valid 3DCORE reconstruction, yielding a total of 61 events, neglecting whether a linked IPS exists in the DONKI database. The sum of all intersection sizes equals 378, which is the total set of ELEvo-predicted arrivals analyzed in this study, as indicated by the bar labeled as ``Category size" in the bottom row of the matrix.  

Beyond the intrinsic subjectivity of ICME identification, temporal variations in catalog quality and consistency are also likely to contribute to the observed discrepancies. In particular, the DONKI catalog has been compiled over more than a decade, during which analysis procedure, available data products, and community understanding of CME-ICME coupling have evolved substantially. These effects can therefore introduce systematic differences between earlier and more recent listings, and should be kept in mind when interpreting agreement rates between catalogs over long time periods.

\subsection{Reconstruction quality}

As discussed above, neither DONKI arrival times nor the ICMECAT should be regarded as an undisputed ground truth. Consequently, many ARCANE detections that do not coincide with ICMECAT entries may in fact correspond to actual ICMEs, and several of these events likely have valid reconstructions that could, in principle, be evaluated. However, a detailed investigation of all possible cases is beyond the scope of the present study. To ensure an unambiguous reference for evaluation, we therefore restrict the following analysis to events for which we have a clear accordance between ARCANE detection with an associated 3DCORE reconstruction and ICMECAT event. Specifically, we combine the two groups highlighted in red from the upset plot in Figure \ref{fig:upset_plot} that contain both an ARCANE detection, a 3DCORE reconstruction and an ICMECAT event, resulting in 61 instances.

Although the available reconstructions and NEXUS outputs would allow for a wide range of possible evaluation strategies, we restrict the analysis to a focused set of metrics that are directly relevant for space weather applications and whose interpretation is straightforward. With $\mathbf{B}(t)$ denoting the observed magnetic field vector and $\hat{\mathbf{B}}(t)$ the corresponding 3DCORE ensemble mean, we consider:

\begin{enumerate}
    \item Root-mean-square error of total field strength
    \[
    \mathrm{RMSE}(B_T)=\sqrt{\frac{1}{N}\sum_{i=1}^{N}\left(|\mathbf{\hat{B}}(t_i)-\mathbf{B}(t_i)|\right)^2}
    \]

    \item Root-mean-square error of the $B_z$ component
    \[
    \mathrm{RMSE}(B_Z)=\sqrt{\frac{1}{N}\sum_{i=1}^{N}\left(|\hat{B}_Z(t_i)-B_Z(t_i)|\right)^2}
    \]

    \item Error in peak total field strength
    \[
    \Delta B_{T,\max} = \max|\hat{\mathbf{B}}(t)|-\max|\mathbf{B}(t)|
    \]

    \item Error in the minimum $B_z$ component
    \[
    \Delta B_{Z,\min} = \min \hat B_Z(t)-\min B_Z(t)
    \]

    \item Timing error of $B_T$ maximum
    \[
    \Delta t_{B_{T,\max}} = t\!\left(\max |\mathbf{\hat B}|\right)-t\!\left(\max |\mathbf{B}|\right)
    \]

    \item Timing error of minimum $B_Z$
    \[
    \Delta t_{B_{Z,\min}} = t\!\left(\min \hat{B}_Z\right)-t\!\left(\min B_Z\right)
    \]

    \item Error in MO start time
    \[
    \Delta t_{\mathrm{start}} = \hat t_{\mathrm{start}}-t_{\mathrm{start}}
    \]

    \item Error in MO end time
    \[
    \Delta t_{\mathrm{end}} = \hat t_{\mathrm{end}}-t_{\mathrm{end}}
    \]
\end{enumerate}

To account for different event durations, timing metrics are additionally normalized by the catalog event duration $\Delta t_{\mathrm{event}}$.

\begin{enumerate}\setcounter{enumi}{8}
        \item Normalized timing error of $B_T$ maximum
    \[
    \frac{\Delta t_{B_{T,\max}}}{\Delta t_{\mathrm{event}}}
    \]

    \item Normalized timing error of minimum $B_Z$
    \[
    \frac{\Delta t_{B_{Z,\min}}}{\Delta t_{\mathrm{event}}}
    \]

    \item Normalized error in MO start time
    \[
    \frac{\Delta t_{\mathrm{start}}}{\Delta t_{\mathrm{event}}}
    \]

    \item Normalized error in MO end time
    \[
    \frac{\Delta t_{\mathrm{end}}}{\Delta t_{\mathrm{event}}}
    \]
\end{enumerate}

\begin{figure}
    \begin{center}
    \includegraphics[width=\textwidth]{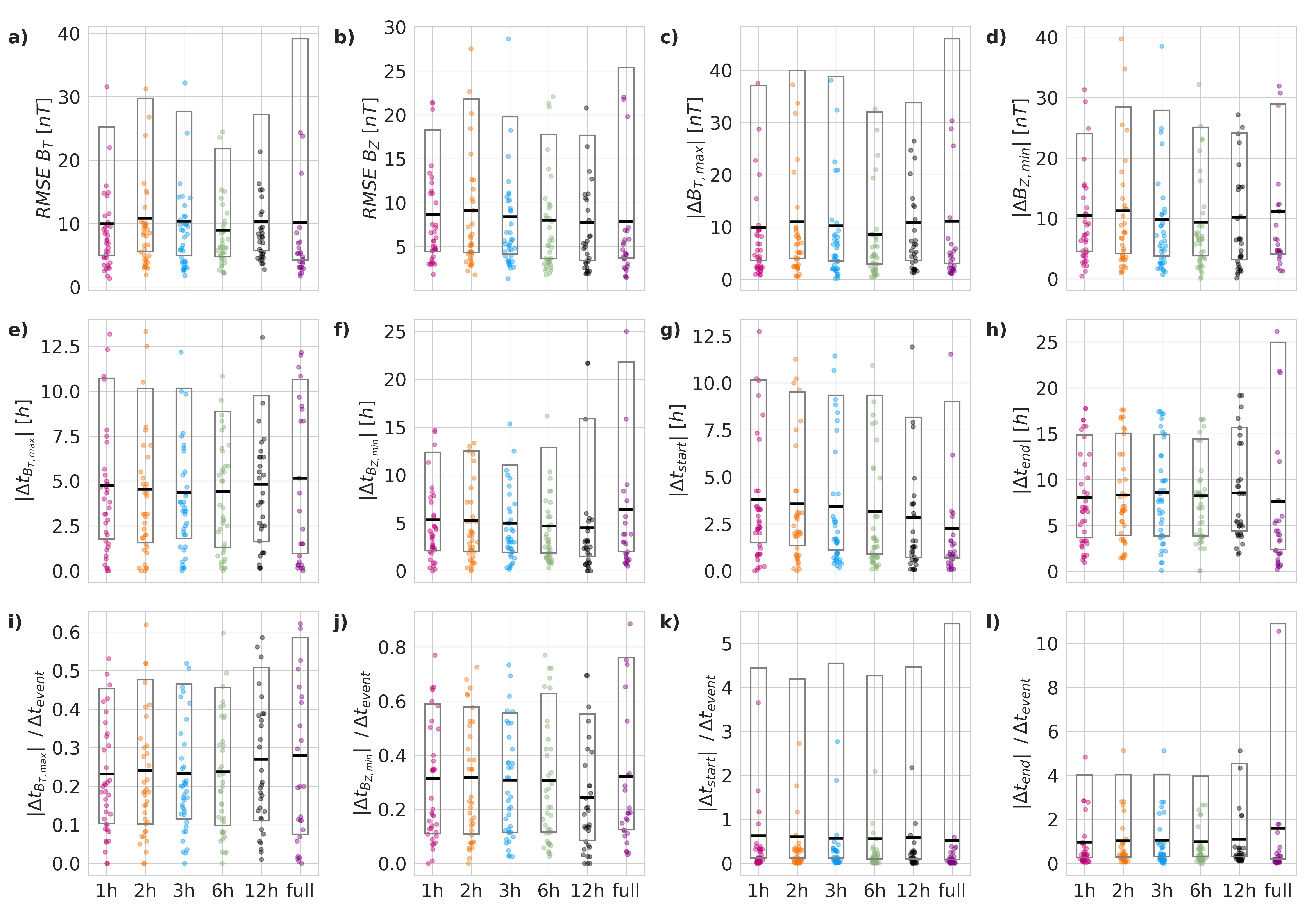}
    \caption{Metrics as a function of observation time. Each panel shows the distribution of the respective evaluation metric for all events and prediction times (1~h, 2~h, 3~h, 6~h, 12~h and for the full event). (a) RMSE of the total magnetic field strength, (b) RMSE of the $B_Z$ component, (c) absolute difference between the maximum magnitude of the magnetic field vector of the actual observations and the 3DCORE reconstruction, (d) absolute difference between the minimum of the $B_Z$ component of the actual observations and the 3DCORE reconstruction, (e) absolute timing error of the maximum of the magnetic field magnitude, (f) absolute timing error of the minimum of the $B_Z$ component, (g) absolute error of the predicted MO start time, (h) absolute error of the predicted MO end time, (i-l) same timing-related metrics as in (e-h), but normalized by the true event duration. Individual dots represent the values for single events (outliers clipped at the 95th percentile for visual clarity). The thick black horizontal line denotes the mean of the full (unclipped) distribution. Gray boxes indicate the asymmetric spread around the mean, computed separately for positive and negative deviations, such that the upper and lower extensions reflect how strongly the metric varies above and below the mean.}
    \label{fig:quality_vs_time_dot}
    \end{center}
\end{figure}

Figure~\ref{fig:quality_vs_time_dot} summarizes the behavior of all metrics as a function of the time elapsed since MO detection (1~h, 2~h, 3~h, 6~h, 12~h, full event). Individual dots correspond to single events, while the mean and asymmetric spread of the full distributions are shown by the black line and gray boxes, respectively. Table~\ref{tab:metrics} lists corresponding numerical summary statistics.

Interestingly, the RMSE of both $B_T$ and $B_Z$ remains relatively stable as more of the event is observed. This indicates that the ensemble reconstructions already capture the overall magnetic field amplitude reasonably well at very early stages of the MO and do not systematically improve simply by fitting to more data. In other words, additional in situ measurements during the event do not automatically translate to lower errors, suggesting that the dominant error source lies in the simplified model or the fact that some events may not exhibit a well-defined flux rope, rather than in lack of observational constraint. Another contributing factor to the limited improvement is that observed extrema in $B_T$ and $B_Z$ are often reached close to the onset of the MO and are therefore already captured during the earliest fitting stages. As a result, the parts of the profile that contribute most strongly to the overall error are constrained early on, while the later evolution typically introduces comparatively smaller deviations.

All metrics exhibit substantial event-to-event variability. This reflects not only the wide diversity of ICME magnetic field signatures, but also supports the view that only a subset of ICMEs is well described by an idealized flux rope model. Events with strong distortions, erosion, interaction, or general deviations from the idealized picture naturally lead to poorer fits, even when sufficient data is available.

A clear systematic trend is seen in the bias of $\Delta B_{T,\max}$ and $\Delta B_{Z,\min}$. For $\Delta B_{T,\max}$, the bias is consistently negative, indicating that the maximum field strength tends to be underestimated, whereas the bias for $\Delta B_{Z,\min}$ is consistently positive, meaning that the minimum $B_Z$ is often overestimated. Together, these biases imply that reconstructed structures are typically ``less severe” than the observations.

The error in the MO end time shows only weak dependence on the fraction of the event observed. This suggests that the assumed mean MO duration of 22~h already provides an estimate that is, on average, as accurate as (and in some cases better than) the duration inferred from the automated detection. This also illustrates the difficulty of accurately identifying the trailing edge of the MO, a boundary that is often less clearly defined than the onset. At least partially, it might also be due to the fact that short CMEs often have less clear signatures and might therefore be missed by the automatic detection. This would lead to a bias towards longer MO durations, explaining why the estimated duration of 22~h might be a good guess for most of the analyzed CMEs.

Furthermore, the full-event reconstructions do not necessarily outperform short-term forecasts based on only a few hours of MO data. This is likely due to two effects. First, the initial parameters and fitting points are not optimized to minimize the evaluation metrics, as typically done in hindsight event analysis, but rely on automatically set fitting points and predefined value ranges. Second, model dependent simplifications dominate over data incompleteness for many events, such that additional data cannot fully compensate for intrinsic model limitations.

Overall, the reconstruction quality with the RMSE of $B_Z < 10$~nT, the RMSE of $B_T < 11$~nT, and the errors on the extrema of $B_Z$ and $B_T$ around $10$~nT across all reconstructions is encouraging given the limited information available at early stages of MO passage and the simplicity of the underlying flux rope model. At the same time, the substantial scatter across events and the systematic biases in amplitudes reinforce the conclusion that many ICMEs deviate significantly from idealized flux rope models and that more sophisticated models and methods are essential for operational use.

\begin{table}
\centering
\begin{tabular}{llcccccc}
\hline
Metric & Stat. & 1h & 2h & 3h & 6h & 12h & full \\
\hline
\multirow{3}{*}{RMSE $B_T$} & Mean $|x|$ & 9.99 & 10.91 & 10.42 & 9.00 & 10.38 & 10.17 \\
 & Bias & 9.99 & 10.91 & 10.42 & 9.00 & 10.38 & 10.17 \\
 & Std & 9.33 & 10.40 & 10.61 & 7.90 & 9.79 & 13.81 \\
\hline
\multirow{3}{*}{RMSE $B_Z$} & Mean $|x|$ & 8.70 & 9.16 & 8.41 & 8.02 & 7.75 & 7.87 \\
 & Bias & 8.70 & 9.16 & 8.41 & 8.02 & 7.75 & 7.87 \\
 & Std & 6.66 & 8.27 & 7.65 & 6.94 & 7.01 & 8.59 \\
\hline
\multirow{3}{*}{$\Delta B_{T,\max}$} & Mean $|x|$ & 9.92 & 11.03 & 10.26 & 8.64 & 10.87 & 11.15 \\
 & Bias & -9.69 & -10.99 & -10.22 & -8.28 & -10.67 & -10.99 \\
 & Std & 14.02 & 14.90 & 14.99 & 12.46 & 14.37 & 18.36 \\
\hline
\multirow{3}{*}{$\Delta B_{Z,\min}$} & Mean $|x|$ & 10.49 & 11.29 & 9.84 & 9.40 & 10.21 & 11.18 \\
 & Bias & 9.92 & 11.05 & 9.80 & 8.74 & 9.81 & 11.18 \\
 & Std & 9.99 & 11.94 & 11.29 & 10.14 & 10.64 & 11.99 \\
\hline
\multirow{3}{*}{$\Delta t_{B_{T,\max}}$} & Mean $|x|$ & 4.77 & 4.55 & 4.37 & 4.41 & 4.82 & 5.16 \\
 & Bias & -1.34 & -0.98 & -0.55 & 0.07 & -0.21 & 1.12 \\
 & Std & 6.38 & 6.21 & 5.92 & 5.89 & 6.40 & 7.10 \\
\hline
\multirow{3}{*}{$\Delta t_{B_{Z,\min}}$} & Mean $|x|$ & 5.35 & 5.26 & 4.98 & 4.70 & 4.51 & 6.41 \\
 & Bias & 2.79 & 1.57 & 2.25 & 2.22 & 2.42 & 4.10 \\
 & Std & 6.77 & 7.09 & 6.35 & 6.71 & 7.36 & 10.18 \\
\hline
\multirow{3}{*}{$\Delta t_{\mathrm{start}}$} & Mean $|x|$ & 3.82 & 3.63 & 3.48 & 3.31 & 3.17 & 2.23 \\
 & Bias & -0.63 & -0.64 & -0.25 & -1.20 & -2.45 & -1.65 \\
 & Std & 5.63 & 5.38 & 5.40 & 5.24 & 4.79 & 3.78 \\
\hline
\multirow{3}{*}{$\Delta t_{\mathrm{end}}$} & Mean $|x|$ & 8.38 & 8.66 & 8.91 & 8.49 & 8.72 & 8.02 \\
 & Bias & 1.83 & 1.87 & 2.26 & 1.51 & 0.54 & 4.24 \\
 & Std & 10.05 & 10.28 & 10.43 & 10.08 & 10.51 & 12.11 \\
\hline
\multirow{3}{*}{$\Delta t_{B_{T,\max}}/\Delta t_{\mathrm{event}}$} & Mean $|x|$ & 0.23 & 0.24 & 0.23 & 0.24 & 0.27 & 0.28 \\
 & Bias & -0.04 & -0.01 & 0.02 & 0.04 & 0.06 & 0.12 \\
 & Std & 0.29 & 0.31 & 0.29 & 0.30 & 0.33 & 0.36 \\
\hline
\multirow{3}{*}{$\Delta t_{B_{Z,\min}}/\Delta t_{\mathrm{event}}$} & Mean $|x|$ & 0.31 & 0.32 & 0.31 & 0.31 & 0.24 & 0.32 \\
 & Bias & 0.07 & -0.00 & 0.05 & 0.06 & 0.10 & 0.17 \\
 & Std & 0.39 & 0.40 & 0.38 & 0.40 & 0.32 & 0.41 \\
\hline
\multirow{3}{*}{$\Delta t_{\mathrm{start}}/\Delta t_{\mathrm{event}}$} & Mean $|x|$ & 0.63 & 0.61 & 0.58 & 0.57 & 0.63 & 0.50 \\
 & Bias & -0.46 & -0.44 & -0.40 & -0.47 & -0.59 & -0.46 \\
 & Std & 1.57 & 1.51 & 1.50 & 1.49 & 1.59 & 1.58 \\
\hline
\multirow{3}{*}{$\Delta t_{\mathrm{end}}/\Delta t_{\mathrm{event}}$} & Mean $|x|$ & 0.95 & 1.00 & 1.03 & 0.96 & 1.08 & 1.56 \\
 & Bias & 0.75 & 0.80 & 0.83 & 0.75 & 0.82 & 1.41 \\
 & Std & 1.56 & 1.64 & 1.63 & 1.58 & 1.82 & 3.95 \\
\hline
\end{tabular}
\caption{Summary statistics for all metrics shown in Figure~\ref{fig:quality_vs_time_dot}. For each metric and observation time, three quantities are listed: the mean of the absolute values (Mean $|x|$), the bias (mean of the signed values $x$), and the standard deviation of the signed values (Std). For the normalized timing metrics, all quantities are additionally divided by the true event duration $\Delta t_{\mathrm{event}}$.}
\label{tab:metrics}
\end{table}

%%%%%%%%%%%%%%%%%%%%%%%%%%%%%%%%%%%%%%%%%%%%%%%
%  Discussion
%%%%%%%%%%%%%%%%%%%%%%%%%%%%%%%%%%%%%%%%%%%%%%%

\section{Discussion}\label{sec:discussion}

A central finding of the evaluation is the weak correspondence between data sources (ICMECAT events, DONKI IPS arrivals, ARCANE detections and ELEvo predictions). This is consistent with the long-recognized subjectivity of ICME definitions and boundary selections. Recent meta-catalog work has highlighted substantial discrepancies across catalogs and reconstruction approaches \citep[e.g.][]{kay_2026_collectioncollationcomparison, kay_2024_collectioncollationcomparison}. In this context, it is neither realistic nor necessarily desirable to optimize the pipeline for perfect agreement with any single catalog.

In addition, our event selection is influenced by operational constraints imposed on the DONKI inputs. To mimic a realistic real-time scenario, we excluded DONKI entries that were submitted more than 8 hours after the CME reached 21.5~\rs, as described in Section~\ref{sec:data}. While this restriction is necessary to ensure that the results presented here could, in principle, have been obtained in real-time, it further reduces the number of usable events by removing cases with delayed or revised analyses. In practice, this excludes all DONKI events prior to the start of our analysis period in 2013 and also removes a subset of events within the years considered. Consequently, some ICMEs with clear in situ signatures may lack a corresponding usable DONKI entry and are therefore not included in our study. This selection effect should be kept in mind when interpreting the reported event statistics.

Operationally, the objective is to identify CMEs with sufficient timeliness to enable downstream modeling and produce actionable estimates of key geoeffective quantities (e.g. the magnitude and timing of southward $B_Z$) with quantified uncertainty. The present results suggest that NEXUS can achieve this for a subset of events. Across the evaluated events, the RMSE of both $B_T$ and $B_Z$ shows only limited systematic improvement with increasing observation time. This implies that, for many events, the dominant error source is not the lack of observational constraint early in the MO, but rather model inadequacy and/or non-flux-rope-like structure. In other words, once the early MO segment has been observed, additional data does not necessarily correct mismatches between an idealized flux rope model and the true event structure. This interpretation is consistent with the event-to-event scatter found across all metrics. 

The limited improvement with increasing observation time may also at least partly result from the automatic choice of fitting points. Early reconstructions place fitting points relatively close together near the start of the MO, which is typically where the MOs magnetic field reaches its extreme values. In contrast, the full-event reconstruction uses equally spaced fitting points across the entire interval. This uniform placement may fail to sufficiently emphasize extrema during the optimization, potentially leading to comparatively poor fits despite the availability of more data. Future versions of NEXUS could address this by adopting adaptive fitting-point strategies that prioritize physically relevant features, while still preserving full automation.

These findings support two practical conclusions. First, the early MO segment appears to contain much of the information that determines the subsequent evolution within the constraints of a simple flux rope model. The fact that early signatures already carry predictive information is also consistent with \citet{reiss_2021_machinelearningpredicting}, who showed that statistics from the sheath and early MO can be used to predict quantities such as minimum $B_Z$ and maximum $B_T$. Second, improving forecast skill for the broader CME population will likely require more realistic (deformable) models or complementary approaches that can represent non-ideal ejecta and interactions, rather than merely extending the fitting window.

For an unambiguous evaluation, we restricted the reconstruction analysis to cases with clear agreement between ARCANE detections and ICMECAT events. This yields a comparatively small set of 61 events and likely introduces a selection bias. ``Clean" and coherent MOs are easier to label consistently, easier to detect, and also the ones a flux rope model can reproduce best. Meta-catalog comparisons indicate that such ``clean" events tend to be the most consistently identified across different catalogs \citep{kay_2026_collectioncollationcomparison}. Consequently, the reconstruction statistics presented here should be interpreted as performance on a favorable subset, rather than as a universal performance claim for all ICMEs.

Beyond the analyses presented here, the archived real-time runs generated by this work enable additional evaluation strategies. For example, one may quantify how reconstructed 3DCORE parameters evolve as additional MO data becomes available, including the stability of inferred properties such as handedness. More application-oriented metrics could also be investigated, such as the duration and integrated strength of southward $B_Z$ or direct proxies for geomagnetic response. In addition, the present study focused on a comparatively small subset of events with clear counterparts in the ICMECAT. Future work should also explore how to best assess reconstruction quality for events without a catalog counterpart, including actual real-time scenarios \citep[e.g.][]{davies_2025_realtimepredictiongeomagneticb}.

Additionally, the boundaries of the ARCANE-detected MO can still differ substantially from the corresponding ICMECAT MO. This matters because flux rope fitting results are highly sensitive to interval selection \citep[e.g.][]{al-haddad_2013_magneticfieldconfiguration, al-haddad_2018_fittingreconstructionthirteen}. Changes in start/end times can lead to different best-fit parameters and thus different inferred global properties. This dependence has been demonstrated in meta-catalog comparisons by \cite{kay_2026_collectioncollationcomparison} and should be considered when interpreting the reconstruction statistics presented here.

At the same time, this bias can be operationally meaningful. If NEXUS preferentially triggers and reconstructs those events for which a flux-rope-based short-term forecast is physically appropriate, then the system may be self-filtering in a beneficial way. Moreover, we showed in \cite{rudisser_2026_arcaneearlydetectioninterplanetary} that high-impact events (characterized by high magnetic field strength and high speed) are detected more consistently than weaker, slower events, suggesting that the subset captured by NEXUS may be skewed toward operationally relevant cases. This effect may be further reinforced by the fact that flux rope MOs are statistically associated with stronger geomagnetic responses than non-flux-rope ejecta \citep[e.g.,][]{zhang_2007_solarinterplanetarysources, yermolaev_2012_geoeffectivenessefficiencycir}. Consequently, a pipeline that preferentially identifies events with clear flux rope signatures may implicitly focus on those CMEs that are more likely to be geoeffective.

It is important to note that not all events contribute equally to the distributions shown at each reconstruction time. Because event start times may be updated during the detection process, not all events necessarily persist long enough to reach later reconstruction intervals, and some reconstructions may fail to converge within the real-time constraints, the number of available reconstructions can differ between the 1~h, 2~h, 3~h, 6~h, 12~h and full-event stages. In the present analysis, metrics were aggregated per reconstruction time without explicitly weighing in this imbalance.

Interacting events and compound structures represent an additional limitation. Some of the most geoeffective disturbances are driven by interacting CMEs (or complex ejecta), as highlighted by recent extreme-event case studies \citep[e.g.][]{scolini_2020_cmecmeinteractionssources, weiler_2025_firstobservationsgeomagnetic}. Such events are unlikely to be well described by a single idealized flux rope and challenge both arrival time prediction, in situ detection, and reconstruction. ELEvo does not model CME-CME interactions, implying that predicted arrival windows may overlap, multiple CMEs may fall into one window, or a single in situ structure may plausibly be associated with multiple solar events. Within the current pipeline, this can lead to ambiguous associations between remote-sensing entries and in situ counterparts and may result in the same in situ ICME being reconstructed multiple times for different ``original" DONKI events. In our evaluation sample, the 61 reconstructions correspond to only 46 unique ICMECAT events. We deliberately did not exclude such duplicate instances, as they represent situations that can realistically occur in an operational real-time setting. However, for the statistical analysis presented here, this implies that some ICMECAT events contribute multiple reconstructions. These reconstructions may differ in their inferred parameters and thus their quality, because the initial conditions derived from the associated DONKI entries (e.g. launch time, propagation direction, or speed) are not necessarily similar. We did not attempt to quantify the effect of such duplicate reconstructions on the reported performance metrics. Assessing how interaction-driven ambiguities and multiple remote-sensing associations affect reconstruction consistency and forecast skill represents an important topic for future work.

From an operational design perspective, the present architecture makes a deliberate trade-off: ELEvo provides an arrival time prior to reduce the search space for in situ detection, and ARCANE then provides the in situ confirmation required to trigger reconstruction. This coupling can reduce unnecessary reconstructions and improves matching between remote-sensing and in situ counterparts, but it also means that errors or biases in the arrival time prior can propagate into missed events. Conversely, substantially widening the arrival window may increase sensitivity but reduce association uniqueness, particularly during periods of high CME activity. The current setup was chosen as a balance between these requirements.

A related operational limitation arises in situations where remote-sensing data are unavailable, for example due to coronagraph outages, which would prevent timely DONKI entries and thus the initialization of ELEvo. In principle, NEXUS could still operate in such cases by continuously applying ARCANE to the in situ data stream and attempting reconstructions for all detected events without an arrival time prior. However, a key strength of the current architecture is that ELEvo constrains the search window, substantially reducing ARCANE false positives and enabling association between in situ detections and their solar counterparts. Removing this prior would likely increase the false-alarm rate and eliminate the linkage to a specific solar eruption. In particular, without an associated DONKI entry, the CME launch time could no longer be treated as a fixed parameter in 3DCORE, and constraints on the initial speed and propagation direction would be lost. A possible workaround would be to extract additional information directly from the in situ plasma measurements. While in the present implementation we use plasma data only to estimate the background solar wind speed prior to the event, the observed CME arrival speed itself could, in principle, be used to infer the original launch speed and thereby restrict the 3DCORE parameter ranges even in the absence of remote-sensing data. Exploring such strategies, and assessing whether they can compensate for the loss of remote-sensing constraints represents an interesting direction for future work but is beyond the scope of the present study.

The iterative ABC-SMC fitting procedure imposes a strict real-time constraint. In this study, the stopping criteria include a hard time limit, ensuring operational feasibility. One potential acceleration strategy would be to initialize each new reconstruction from the refined parameter space of the previous iteration. While this could reduce convergence time, it also risks propagating early-stage fitting errors and reducing exploration of alternative solutions. Given that the current setup did not exhibit severe time bottlenecks when starting from scratch under the imposed limits, we opted for this more robust approach. 

A practical challenge for real-time reconstruction is that the MO end time is unknown while the event is ongoing. Here, we used a fixed estimated duration ($d_e=22$ h) motivated by catalog statistics. Interestingly, the MO end-time error shows only weak dependence on observation time, suggesting that this simple assumption is often competitive with (or at least not dramatically worse than) what can be inferred from automatic detections. Predicting duration dynamically (e.g. through analyzing shock and sheath properties, as well as early MO signatures, or via model-based constraints that link expansion and transit speed to expected size and duration at 1~au) still remains an interesting direction for future studies and could particularly benefit short-duration MOs.

Finally, while the present pipeline is framed around L1 in situ observations, the same logic could be applied to monitors located upstream of L1 (``sub-L1"), thereby increasing lead time for geomagnetic effect forecasting \citep[e.g.][]{palmerio_2025_monitoringsolarwind,weiler_2025_firstobservationsgeomagnetic, davies_2025_realtimepredictiongeomagneticb}. Additionally, the availability of real-time multipoint measurements at sufficiently separated locations, such as spacecraft on distant retrograde orbits that intersect the Sun-Earth line closer to the Sun, could provide the additional constraints to make deformable flux rope models viable in an operational setting \cite[e.g.][]{st.cyr_2000_spaceweatherdiamond, lugaz_2024_neednearearthmultispacecraft, cicalo_2025_missionanalysishenon}. To apply the NEXUS pipeline to such a ``sub-L1" monitor, two requirements would need to be addressed: first, adapting detection and reconstruction to potentially different data and instrument characteristics, and second, forward-modeling the reconstructed structure from the sub-L1 location to L1. This extension is conceptually straightforward within the modular architecture, but would require thorough validation. Another natural extension would be to further couple the short-term forecast with models that directly predict geomagnetic indices, such as Dst or SYM-H, from in situ data \citep[e.g.][]{temerin_2006_dstmodel19952002}. Such a coupling would provide a more direct link between reconstruction quality and geomagnetic effect.

%%%%%%%%%%%%%%%%%%%%%%%%%%%%%%%%%%%%%%%%%%%%%%%
%  Summary
%%%%%%%%%%%%%%%%%%%%%%%%%%%%%%%%%%%%%%%%%%%%%%%

\section{Summary}\label{sec:summary}

In this work, we developed and evaluated the fully automated pipeline ``NEXUS" for real-time CME forecasting that combines remote-sensing-based arrival time prediction, in situ early detection and iterative short-term forecasting of the CME magnetic field structure. NEXUS links CME entries from the CCMC DONKI system with heliospheric propagation modeling using ELEvo, a drag-based model, automatic in situ detection using the deep-learning model ARCANE, and flux rope reconstruction and short-term forecasting using the semi-empirical flux rope model 3DCORE. Importantly, the system operates without human intervention beyond the initial remote-sensing analysis that feeds into DONKI, and all components were evaluated using archived real-time data products rather than curated post-event datasets.

Applying NEXUS to 3870 DONKI-listed CMEs between 2013 and 2025, we demonstrate that automated short-term forecasting of CME magnetic field structure at L1 is feasible for a subset of events. For 61 events with clear correspondence between ARCANE detections and the ICMECAT, which is used as ground truth, short-term forecasts based on only the first hours of MO observations achieve performance comparable to full-event reconstructions under the same automated configuration. Typical errors are on the order of ~5 hours in the timing of key magnetic extrema and ~10~nT in magnetic field strength metrics, with only limited systematic improvement as progressively larger fractions of the MO are observed. The main conclusions of this study are:

\begin{enumerate}
    \item The correspondence between ELEvo predictions, DONKI arrivals, ICMECAT events, and automated detections is limited, underscoring that catalog-based ``ground truth" is inherently uncertain. This motivates evaluation strategies centered on operational utility rather than perfect catalog agreement.
    \item For events with coherent MO signatures that are well described by an idealized flux rope, the pipeline can generate meaningful short-term forecasts early during MO passage. However, substantial event-to-event scatter and systematic underestimation of extrema indicate that many ICMEs significantly deviate from the simplified model employed here.
    \item Forecast quality does not necessarily improve drastically with longer observed fractions of the MO, suggesting that additional in situ data often does not overcome model limitations. Improving forecast skill will therefore require more flexible models and additional observational constraints, that enable their use.
    \item Simple assumptions (such as a fixed estimated MO duration) can be surprisingly competitive, though dynamic duration prediction remains an important future improvement.
    \item The subset of events that are consistently detected and reconstructible is likely biased. Quantifying this selection effect and its operational implications is essential for interpreting performance and designing robust triggers.
\end{enumerate}

Overall, the results demonstrate that real-time, automated short-term forecasting of CME in situ profiles is in theory feasible within a modular pipeline architecture, while also highlighting that uncertainty enters at every stage: from remote-sensing parameter estimation and arrival time prediction, through subjective event definitions, to the limited representational capacity of simplified flux rope models. Additionally, it further supports the theory that only a subset of CMEs exhibit a clean flux rope structure at 1~au \citep[e.g.][]{richardson_2004_fractioninterplanetarycoronal, kilpua_2013_relationshipinterplanetarycoronal}. Future work should focus on extending the pipeline to upstream monitors to increase lead time, improving performance for complex or non-flux-rope CMEs through more flexible models, constrained through additional observations. 

Beyond demonstrating the feasibility of such a pipeline, this study provides a first quantitative evaluation of short-term magnetic field forecasting for those events with a well-defined counterpart in the ICMECAT catalog. For this subset, forecast skill was assessed using a number of statistical metrics. Future work could include detailed event-by-event analysis to better understand which characteristics determine reconstruction success and failure, or the use of additional performance metrics that measure other quantities particularly relevant for space weather, such as integrated southward $B_Z$ or event duration. Equally important is the development of evaluation strategies for events that lack an unambiguous catalog counterpart, as many such events might still be operationally relevant. Finally, coupling the short-term in situ forecasts to geomagnetic response models would provide a direct means of quantifying how reconstruction quality affects our ability in predicting their downstream effects at Earth.

Ultimately, this work establishes a proof of concept for fully autonomous real-time short-term forecasting of CME magnetic field structure based on operational data streams. As such, this study provides both a benchmark for future developments and a concrete step toward operational systems that do not only predict when a CME arrives, but also how severe its geomagnetic effect is likely to be.

%%%%%%%%%%%%%%%%%%%%%%%%%%%%%%%%%%%%%%%%%%%%%%%
%  ACKNOWLEDEGMENTS
%%%%%%%%%%%%%%%%%%%%%%%%%%%%%%%%%%%%%%%%%%%%%%%

\begin{acknowledgments}    
    H.T.~R., E.E.~D., U.V.~A., C.~M. and E.~W. are supported by ERC grant (HELIO4CAST, 10.3030/101042188). Funded by the European Union. Views and opinions expressed are however those of the author(s) only and do not necessarily reflect those of the European Union or the European Research Council Executive Agency. Neither the European Union nor the granting authority can be held responsible for them.
    
    This research was funded in whole or in part by the Austrian Science Fund (FWF) [10.55776/P36093; 10.55776/P34437]. For open access purposes, the author has applied a CC BY public copyright license to any author-accepted manuscript version arising from this submission.
    
    The research leading to these results is part of ONERA Forecasting Ionosphere and Radiation belts Short Time Scale disturbances with extended horizon (FIRSTS) internal project.
    
    We have benefited from the availability of the NOAA RTSW data and thus would like to thank the instrument teams and data archives for their data distribution efforts. 

    We acknowledge the Community Coordinated Modeling Center (CCMC) at Goddard Space Flight Center for the use of the DONKI system, \url{https://kauai.ccmc.gsfc.nasa.gov/DONKI/}.

    During the preparation of this work, the main author used a large language model \citep{chatgpt2026} to partly assist with improving grammar, language, and readability of the manuscript. The tool was not used for data analysis, result generation, or drawing scientific conclusions. All scientific content, interpretations, and conclusions are solely those of the authors, who take full responsibility for the content of the published article.

\end{acknowledgments}

%%%%%%%%%%%%%%%%%%%%%%%%%%%%%%%%%%%%%%%%%%%%%%%
%  BIBLIOGRAPHY
%%%%%%%%%%%%%%%%%%%%%%%%%%%%%%%%%%%%%%%%%%%%%%%

\bibliography{bibliography}{}
\bibliographystyle{aasjournalv7}

\end{document}